\documentclass{aa}
\usepackage{graphicx}

\begin{document}

\title{X-ray emission from the blazar \object{AO 0235+16}:
the XMM-Newton and Chandra point of view\thanks{Based on observations obtained with XMM-Newton, 
an ESA science mission with instruments and contributions directly funded by ESA Member States and NASA,
and with the NASA Chandra X-Ray Observatory}}


\author{C.~M.~Raiteri \inst{1}
\and M.~Villata \inst{1}
\and M.~Kadler \inst{2,3}
\and T.P.~Krichbaum \inst{2}
\and M.~B\"ottcher \inst{4}
\and L.~Fuhrmann \inst{1,5}
\and M.~Orio \inst{1}
}

\offprints{C.\ M.\ Raiteri, \email{raiteri@to.astro.it}}

\institute{INAF, Osservatorio Astronomico di Torino, Via Osservatorio 20,
     10025 Pino Torinese (TO), Italy
\and Max-Planck-Institut f\"ur Radioastronomie, Auf dem H\"ugel 69,
     53121 Bonn, Germany
\and Radioastronomisches Institut der Universit\"at Bonn, Auf dem H\"ugel 71, 53121 Bonn, Germany
\and Dept.\ of Physics and Astronomy, Clippinger 339, Ohio University,
     Athens, OH 45701-2979, USA
\and Osservatorio Astronomico, Universit\`a di Perugia, Via B.\ Bonfigli,
     06126 Perugia, Italy
}

\date{Received; Accepted;}

\titlerunning{X-ray emission from AO 0235+16: XMM-Newton and Chandra}

\authorrunning{C.\ M.\ Raiteri et al.}

\abstract{In this paper we analyse five observations of the BL Lac object \object{AO 0235+16} performed
with the Chandra and XMM-Newton satellites during the years 2000--2005.
In the February 2002 observation the source is found in a bright state
and presents a steep X-ray spectrum, while in all the other epochs it is faint and the spectrum is hard.
The soft X-ray spectrum appears to be strongly absorbed, likely by the intervening system at $z=0.524$, which
also absorbs the optical--UV radiation.
We find that models that consider spectral curvature are superior to single power law ones in fitting
the X-ray spectrum.
In particular, we favour a double power law model,
which agrees with the assumption of a superposition of two different components in the X-ray domain.
Both in the Chandra and in one of the XMM-Newton observations,
a tentative detection of the redshifted Fe K$\alpha$ emission line may suggest its origin
from the inner part of an accretion disc.
Thermal emission from this accretion disc might explain the UV--soft-X-ray bump
that appears in the spectral energy distributions,
when the X-ray spectra are complemented with the optical--UV
data from the Optical Monitor onboard XMM-Newton. 
More likely, the bump can be interpreted in terms of
an additional synchrotron component emitted from an inner region of the jet with respect 
to that where the lower-energy emission comes from.
An inspection of the X-ray light curves reveals that intraday variability occurs only when the
source is in a bright state.
}

\maketitle

\section{Introduction}

The BL Lac object AO 0235+16 ($z=0.94$) is known to be an intriguing source,
exhibiting a number of extreme properties. It has shown large amplitude variability
on long time scales (months--years) and noticeable intraday variability (IDV) over a wide range
of the electromagnetic spectrum, from the radio to the optical band (Takalo et al.\ \cite{tak92};
Heidt \& Wagner \cite{hei96}; Noble \& Miller \cite{nob96}; Romero et al.\ \cite{rom97};
Kraus et al.\ \cite{kra99}; Romero et al.\ \cite{rom00}; Raiteri et al.\ \cite{rai01}, \cite{rai05}).
These observations support the idea that the radiation from this object is strongly beamed
towards us, with Doppler factors reaching up to $\delta \sim 100$  
(Fujisawa et al.\ \cite{fuj99}; Frey et al.\ \cite{fre00}; Jorstad et al.\ \cite{jor01}; Kraus et al.\ \cite{kra03}).

The interpretation of optical
variability is made difficult by the presence of a complex environment
(Yanny et al.\ \cite{yan89}; Burbidge et al.\ \cite{bur96}; Nilsson et al.\ \cite{nil96}).
A couple of objects at redshift $z=0.524$ are very close to the line of sight: one is an
AGN (named ELISA by Raiteri et al.\ \cite{rai05}) located 2 arcsec south of the source,
affecting its optical--UV photometry.
This object is probably interacting with the other close object, a galaxy
lying 1.3 arcsec east of AO 0235+16, and this intervening system strongly absorbs
the IR-to-soft-X-ray radiation of the blazar.
Absorption lines due to this damped Ly$\alpha$ (DLA)
intervening system are seen in the optical and UV spectra of AO 0235+16.
In particular, a fit to the Ly$\alpha$ absorption line profile of the Hubble Space Telescope (HST) spectrum
taken in February 1998 led to an estimate of the hydrogen column density at $z=0.524$, which is
$N_{\ion{H}{i}} = 4.5 \pm 0.4 \times 10^{21} \, \rm cm^{-2}$ according to Turnshek et al.\ (\cite{tur03}), and
$N_{\ion{H}{i}} = 5 \pm 1 \times 10^{21} \, \rm cm^{-2}$ according to Junkkarinen et al.\ (\cite{jun04}).

Moreover, objects in the same intervening system could act as microlenses for the source emission, so that
an interpretation of at least some episodes of variability in terms of extrinsic processes cannot be ruled out
(Stickel et al.\ \cite{sti88}; Takalo et al.\ \cite{tak98}; Webb et al.\ \cite{web00};
but see also the critical discussion by Kayser \cite{kay88}).

AO 0235+16 has recently been the target of a huge multiwavelength
observing effort led by the Whole Earth Blazar Telescope\footnote{{\tt http://www.to.astro.it/blazars/webt/}}
(WEBT; e.g.\ Villata et al.\ \cite{vil00}, \cite{vil02},
\cite{vil04a}, \cite{vil04b}) in the observing seasons 2003--2004 and 2004--2005.
Three XMM-Newton pointings were awarded to get information
on the high-energy emission of the source.
The results of the radio to optical monitoring during the first season of the campaign, together with
an analysis of the XMM-Newton pointing of January 2004, have already been published by
Raiteri et al.\ (\cite{rai05}).
Here we present the results of the XMM-Newton observations performed in August 2004 and January 2005,
during the second observing season of the WEBT campaign.
The other multiwavelength data collected by the WEBT
in this period will be published elsewhere (Raiteri et al.\ \cite{rai06}).
We also analyse an archival XMM-Newton pointing performed in February 2002
and re-analyse the Chandra data taken in August 2000 (Turnshek et al.\ \cite{tur03})
and those from the XMM-Newton pointing of January 2004.
The aim is to better understand both the X-ray properties of this enigmatic source through a
homogeneous analysis of the most recent data and their relationship with the
lower-energy emission.
In particular, we investigate the existence of a spectral curvature in the X-ray energy range 
as well as the presence of a bump in the UV--soft-X-ray region of the SED.

This paper is organized as follows:
in Section 2 we review past observations of AO 0235+16 by X-ray satellites.
A detailed analysis of the X-ray data acquired by Chandra in August 2000 and
by XMM-Newton in 2002--2005 is performed in Section 3, where
the possibility that the fluorescent Fe K$\alpha$ line was detected in some of the above spectra
is also discussed.
In Section 4 the X-ray spectra are complemented by the optical and UV data from the
Optical Monitor onboard XMM-Newton and the resulting UV--soft-X-ray bump is discussed in Section 5.
The X-ray light curves corresponding to the above pointings are presented in Section 6.
Conclusions are drawn in Section 7.

\section{Previous X-ray observations by other satellites}

\begin{table*}
\centering
\caption{Results from previous X-ray observations of AO 0235+16 by other satellites. 
When two numbers appear in the $N_{\rm H}$
column, the first refers to the Galactic absorption, 
the second to absorption by the $z=0.524$ intervening system.
$F_{1 \, \rm keV}$ is the unabsorbed 1 keV flux density}
\begin{tabular}{l c c c c c c}
\hline
Satellite & Observation  & Energy range &$N_{\rm H}$ & $\Gamma$               & $F_{1 \, \rm keV}$  &Reference$^{a}$\\
          & date         & (keV)        &($10^{21} \, \rm cm^{-2}$)&                        & ($\mu \rm Jy$) &\\
\hline
Einstein  &1979--1981    & 0.1--3.5     &$> \rm 0.76$               & 3.25$^{+4.9}_{-2.1}$   & 1.51        &(1)\\
EXOSAT    &Aug.\ 1, 1984 & 0.1--10      &$8^{+9}_{-7}$& 1.75$^{+0.80}_{-0.65}$ & 1.44$^{+4.64}_{-1.33}$    &(2)\\
ROSAT     &Jul.\ 21 -- Aug.\ 15, 1993 & 0.1--2.1 & 0.76, 3.8$^{+1.2}_{-1.1}$ & 2.67$\pm 0.32$  & 1.24      &(3)\\
ROSAT     &Jul.\ 26, 1993   & 0.1--2.0  & 2.46$^{+1.67}_{-1.22}$ & 2.59$^{+0.97}_{-0.76}$ & 1.15           &(4)\\
ROSAT     &Aug.\ 15, 1993   & 0.1--2.0  & 2.61$^{+2.28}_{-1.58}$ & 2.54$^{+0.88}_{-0.97}$ & 0.41           &(4)\\
ASCA      &Feb. 4--19, 1994 &  0.5--10  & 0.76, 4.6$\pm 1.3$         & 1.96$\pm 0.09$ & 0.3                &(3)\\
RXTE      &Nov.\ 3--6, 1997 &   2--60   & -                      & 1.41$\pm 0.20$ & (0.14$\pm 0.02$)$^{b}$ &(5)\\
ASCA      &Feb.\ 11--12, 1998& 0.5--10  & 0.76, 4.6$\pm 1.3$         & 1.72$\pm 0.07$ &  0.278             &(6)\\
BeppoSAX  &Jan.\ 28, 1999   & 0.1--10   & 2.8                    & 1.96$^{+0.27}_{-0.26}$ & 0.20$^{+0.08}_{-0.06}$ &(7)\\
\hline
\multicolumn{7}{l}{$^{a}$ (1) Worral \& Wilkes (\cite{wor90});
(2) Ghosh \& Soundararajaperumal (\cite{gho95});
(3) Madejski et al.\ (\cite{mad96}); }\\
\multicolumn{7}{l}{(4) Comastri et al.\ (\cite{com97}); (5) Webb et al.\ (\cite{web00});
(6) Junkkarinen et al.\ (\cite{jun04}); (7) Padovani et al.\ (\cite{pad04})}\\
\multicolumn{7}{l}{$^{b}$ $F_{2 \, \rm keV}$}\\
\label{past}
\end{tabular}
\end{table*}

In this section we briefly review the results derived from past X-ray observations of AO 0235+16.
They are summarized in Table \ref{past} in terms of the hydrogen
column density $N_{\rm H}$ adopted to model the X-ray absorption,
the photon index $\Gamma$\footnote{The photon index $\Gamma$ is related to the energy index
$\alpha$ through $\Gamma=\alpha+1$, recalling that the specific flux at frequency $\nu$ is
$F_\nu \propto \nu^{-\alpha}$}, and the unabsorbed 1 keV flux density.

Worral \& Wilkes (\cite{wor90}) analysed the data taken by the Einstein Observatory
imaging proportional counter (IPC) in 1979--1981. They fit the data with a power law
with free Galactic absorption and 
noticed that an $N_{\rm H}$ value larger than the typical one for Galactic absorption
was required to obtain a good fit.

EXOSAT observed the source in 1983--1986. Data taken during the pointing of August 1, 1984 were
analysed by Ghosh \& Soundararajaperumal (\cite{gho95}) in terms of a simple power law plus absorption;
the highly uncertain value of $N_{\rm H}$ was compatible with the Galactic one.

Madejski et al.\ (\cite{mad96}) presented data taken by ROSAT in summer 1993 and data obtained
by ASCA in February 1994. They first fit the X-ray spectrum with a power law plus free
absorption model, and this led to an extra-absorption. Then, a model
with fixed Galactic absorption plus free absorption at $z=0.524$ was adopted.
The ROSAT data showed AO 0235+16 in a bright and active state, revealing flux variations
up to $\sim 80$\% on a 3 day timescale.
On the contrary, no short-term variability was detected by ASCA, which observed the source
in a much fainter state.
The comparison between the ROSAT and ASCA results revealed spectral variability on the long term,
which was interpreted by Madejski et al.\ (\cite{mad96}) as a dominance of the steep
tail of the synchrotron component when the source is bright, and a
prevalence of the fainter and harder Compton component when it is faint.
The authors ruled out the possibility that a significant Fe K line was present at either
$z=0.524$ or $z=0.94$, with 90\% confidence limits on the line equivalent width of $\sim 60$ 
eV and $\sim 100$ eV, respectively.

Because of the flux variability detected in the ROSAT data, Comastri et al.\ (\cite{com97})
analysed the ROSAT observations of summer 1993 by separating those with low count
rates from those with high count rates. Indeed, they reported a factor $\sim 3$ difference in the flux,
but a more or less constant spectral index.

The source was observed by RXTE in November 1997, during the rising phase of a big 
radio--optical outburst (Webb et al.\ \cite{web00}; Raiteri et al.\ \cite{rai01}). 
Webb et al.\ (\cite{web00})
analysed the X-ray data with a power law model and reported a very low X-ray flux (see Table \ref{past}).
It was suggested that the X-ray response to the low-energy flux increase may be delayed or insignificant.

ASCA pointed again at the source on February 11, 1998; the results were published by Junkkarinen et al.\ (\cite{jun04}).
The spectrum was harder with respect to the previous ASCA pointing, 4 years earlier, and
the flux in the 2--10 keV energy range was 3.4 times higher.

Data acquired by BeppoSAX on January 28, 1999 were reported by Padovani et al.\ (\cite{pad04}).
They found extra-absorption and both spectral index and flux consistent with the
values derived by Madejski et al.\ (\cite{mad96}) for the ASCA data. The problem
of a possible contribution to the X-ray flux by the southern AGN was also discussed, 
leading to the conclusion that it should be roughly a
factor 10 smaller than that observed for AO 0235+16.

We notice that Table \ref{past} can help to trace both the flux and spectral X-ray variability
of the source,
but it is not exhaustive, since in most of the above mentioned papers more than one model
was applied and discussed. Yet, it is not possible to find a common model for all cases, and
the comparison among results from different observations suffers
from inhomogeneous data analysis. It is to avoid such a difficulty that in this paper we have analysed 
or re-analysed the last five X-ray observations using the same data treatment\footnote{A uniform 
re-analysis of the archival ASCA and BeppoSAX data is given in Kadler (\cite{kad05b})}.

\section{XMM-Newton and Chandra X-ray spectra}

In this section we present the X-ray spectra of AO 0235+16 taken by XMM-Newton in August 2, 2004 and January 28, 2005,
during the second observing season of the WEBT campaign on this source.
The results will be compared with those obtained by re-analysing
the XMM-Newton observation performed in January 18--19, 2004, during the first season of the WEBT campaign
(Raiteri et al.\ \cite{rai05}), and with those derived from the analysis of archival XMM-Newton spectra taken
in February 10, 2002. Moreover, we also re-analyse the data of a Chandra pointing on August 20--21, 2000,
which have been already presented by Turnshek et al.\ (\cite{tur03}).
Table \ref{counts} reports the date, the UT of start and end as well as the duration of the exposure
for each pointing and for each instrument.

\begin{table}
\centering
\caption{Log of the Chandra and XMM-Newton observations}
\begin{tabular}{l c c c}
\hline
Detector & Start       & End       & Duration\\
         & (UT)        & (UT)      & (s)     \\
\hline
\hline
\multicolumn{4}{c}{Chandra: August 2000}\\
\multicolumn{4}{c}{PI: D.\ A.\ Turnshek}\\
\hline
ACIS-S   & Aug.\ 20, 23:39:52 & Aug.\ 21, 08:43:05 & 30625\\
\hline
\hline
\multicolumn{4}{c}{XMM-Newton: February 2002}\\
\multicolumn{4}{c}{PI: M. Watson}\\
\hline
MOS1     & Feb.\ 10, 04:27:48 & Feb.\ 10, 09:46:33 & 19125\\
MOS2     & Feb.\ 10, 04:27:47 & Feb.\ 10, 09:46:33 & 19126\\
pn       & Feb.\ 10, 05:01:03 & Feb.\ 10, 09:41:50 & 16847\\
\hline
\hline
\multicolumn{4}{c}{XMM-Newton: January 2004}\\
\multicolumn{4}{c}{PI: C.\ M.\ Raiteri}\\
\hline
MOS1     & Jan.\ 18, 19:07:18 & Jan.\ 19, 03:25:01 & 29863 \\
MOS2     & Jan.\ 18, 19:07:18 & Jan.\ 19, 03:25:06 & 29868 \\
pn       & Jan.\ 18, 19:12:28 & Jan.\ 19, 03:26:59 & 29671 \\
\hline
\hline
\multicolumn{4}{c}{XMM-Newton: August 2004}\\
\multicolumn{4}{c}{PI: C.\ M.\ Raiteri}\\
\hline
MOS1     & Aug.\ 02, 00:42:16 & Aug.\ 02, 03:56:35 & 11659 \\
MOS2     & Aug.\ 02, 00:42:14 & Aug.\ 02, 03:56:40 & 11666 \\
pn       & Aug.\ 02, 00:47:24 & Aug.\ 02, 03:58:35 & 11471 \\
\hline
\hline
\multicolumn{4}{c}{XMM-Newton: January 2005}\\
\multicolumn{4}{c}{PI C.\ M.\ Raiteri}\\
\hline
MOS1     & Jan.\ 28, 14:43:36 & Jan.\ 28, 19:17:57 & 16461 \\
MOS2     & Jan.\ 28, 14:43:40 & Jan.\ 28, 19:18:03 & 16463 \\
pn       & Jan.\ 28, 14:48:46 & Jan.\ 28, 19:19:56 & 16270 \\
\hline
\label{counts}
\end{tabular}
\end{table}

\subsection{Data reduction}

Chandra observed AO 0235+16 on August 20--21, 2000 with ACIS-S.
The image, which has a space resolution of 0.492 arcsec/px,
does not show any emission 2 arcsec south of the source,
thus ruling out an appreciable contribution from the southern AGN, named ELISA.

Data were reduced with the CIAO package version 3.0.2.
We extracted the source spectrum from a circular region of 3 arcsec size around the source
and the background spectrum in a surrounding annulus of radii 5 and 10 arcsec.
Only the energy channels from 0.4 to 8.0 keV were considered.

The European Photon Imaging Camera (EPIC) onboard XMM-Newton
has three detectors: MOS1, MOS2, and pn;
they all point at the same object at the same time.
All XMM-Newton observations analysed in this paper but the one of February 10, 2002,
were performed in small-frame, medium-filter configuration
in order to avoid possible problems of pile-up\footnote{Photon pile-up occurs when more 
than one X-ray photon arrives in one camera pixel or in adjacent pixels before it is read out; 
it can affect both the source point spread function and the spectral response of the instrument.}, 
out-of-time events, and degradation.

The data were reduced with the Science Analysis System (SAS) software version 6.0
\footnote{When extracting the spectrum of both the
source and the background, the strings ``{\tt FLAG==0}" and ``{\tt PATTERN$<$=4}" were 
included in the selection expression for the
{\tt evselect} task of the SAS software for all the three EPIC detectors to provide
conservative event selection criteria.}.
Source spectra were extracted from a circular region centered on the source;
the aperture radius was 30--35 arcsec in all cases but for the February 2002 observations,
where it was 100 arcsec.
Background spectra were derived from source-free regions on the same CCD of the source.
We considered only channels with energy greater than 0.2--0.3 keV and lower than
10--12 keV, depending on the instrument (MOS having stricter limits than pn) and data quality.

In the February 10, 2002 observation the source was detected in a high X-ray state.
As the observation was performed using a full-frame, thin-filter configuration,
we also checked the possible presence of pile-up with the
SAS task {\tt epatplot}. We found a small excess of double events
in the $\sim 1.5$--5 keV energy range in the pn observation,
which was removed by extracting the source spectrum in an annulus with an inner radius of 5 arcsec.
However, the pn spectrum appears somewhat distorted below $\sim 0.5$ keV,
hence for the spectral analysis of this observation we used the data from MOS1 and MOS2 only.

After this preliminary data reduction with the specific packages,
each source spectrum was associated with its response matrix file, ancillary response file,
and background spectrum, as well as binned in order to have a minimum of 25 counts in each bin.
This was accomplished with the task {\tt grppha} of the {\tt FTOOL} package.
These ``grouped" spectra were then analysed with the {\tt Xspec} package, version 11.3.0.
For each epoch, only the reliable energy channels were selected, discarding measurements 
affected by large uncertainties.
For all the XMM-Newton observations, data from the different EPIC detectors are fitted simultaneously, i.e.\
the same model is applied to the MOS1, MOS2, and pn spectra to increase the statistics.

\begin{figure}
   \resizebox{\hsize}{!}{\includegraphics[angle=-90]{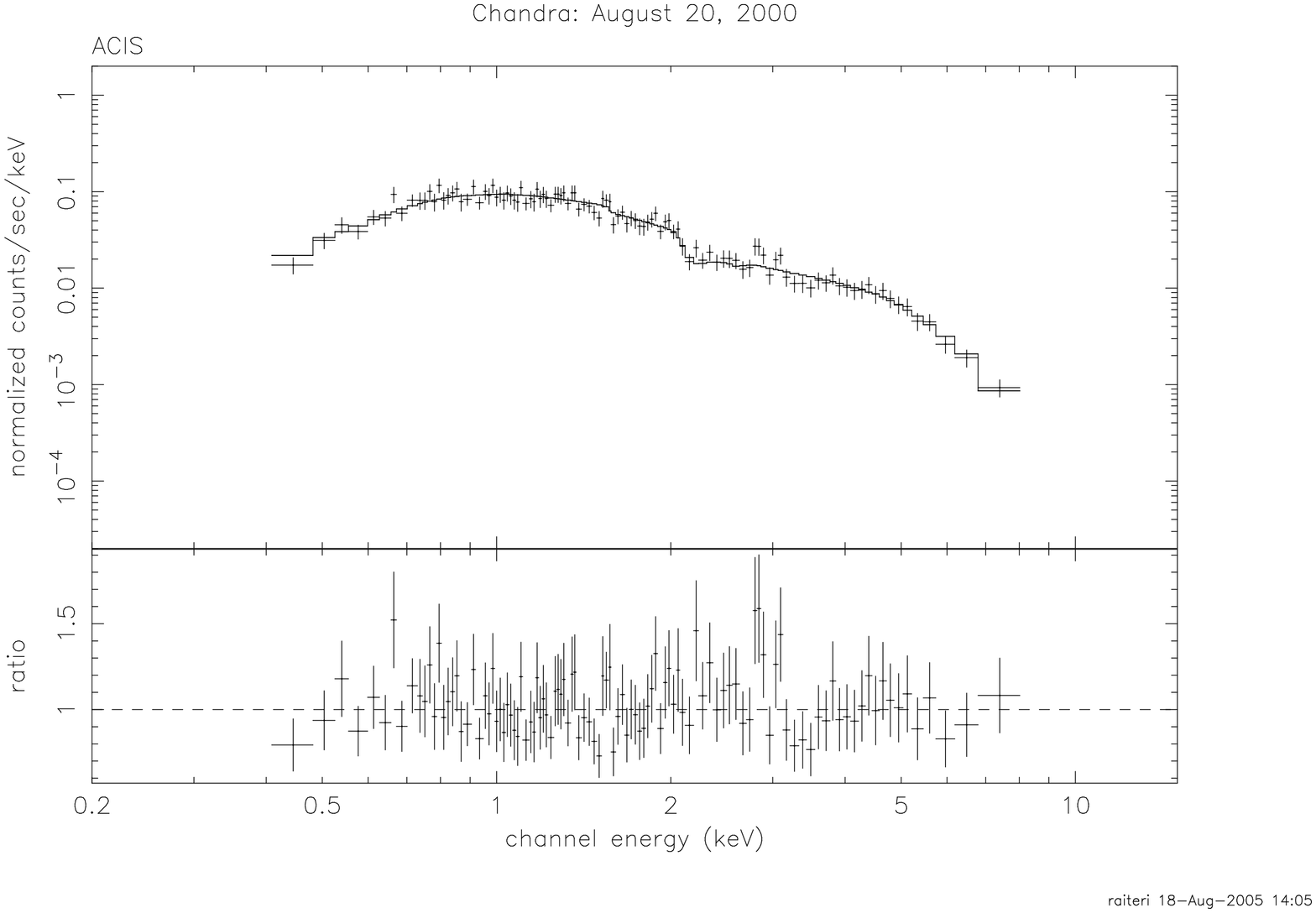}}
   \caption{The X-ray spectrum of AO 0235+16 detected by Chandra on August 20--21, 2000
   fitted by a double power law model with fixed Galactic and $z=0.524$ DLA system absorptions;
   the lower panel shows the ratio of the data to the model fit}
   \label{Chandra}
   \end{figure}

\begin{figure}
   \resizebox{\hsize}{!}{\includegraphics[angle=-90]{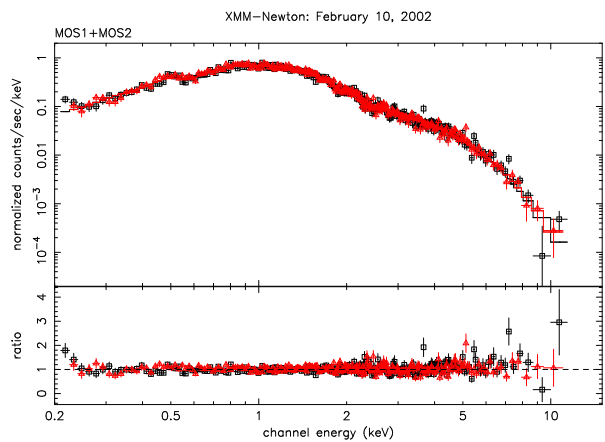}}
   \caption{The X-ray spectrum of AO 0235+16 detected by XMM-Newton on February 10, 2002
   fitted by a double power law model with fixed Galactic and $z=0.524$ DLA system absorptions;
   the lower panel shows the ratio of the data to the model fit}
   \label{XMM0}
   \end{figure}

\begin{figure}
   \resizebox{\hsize}{!}{\includegraphics[angle=-90]{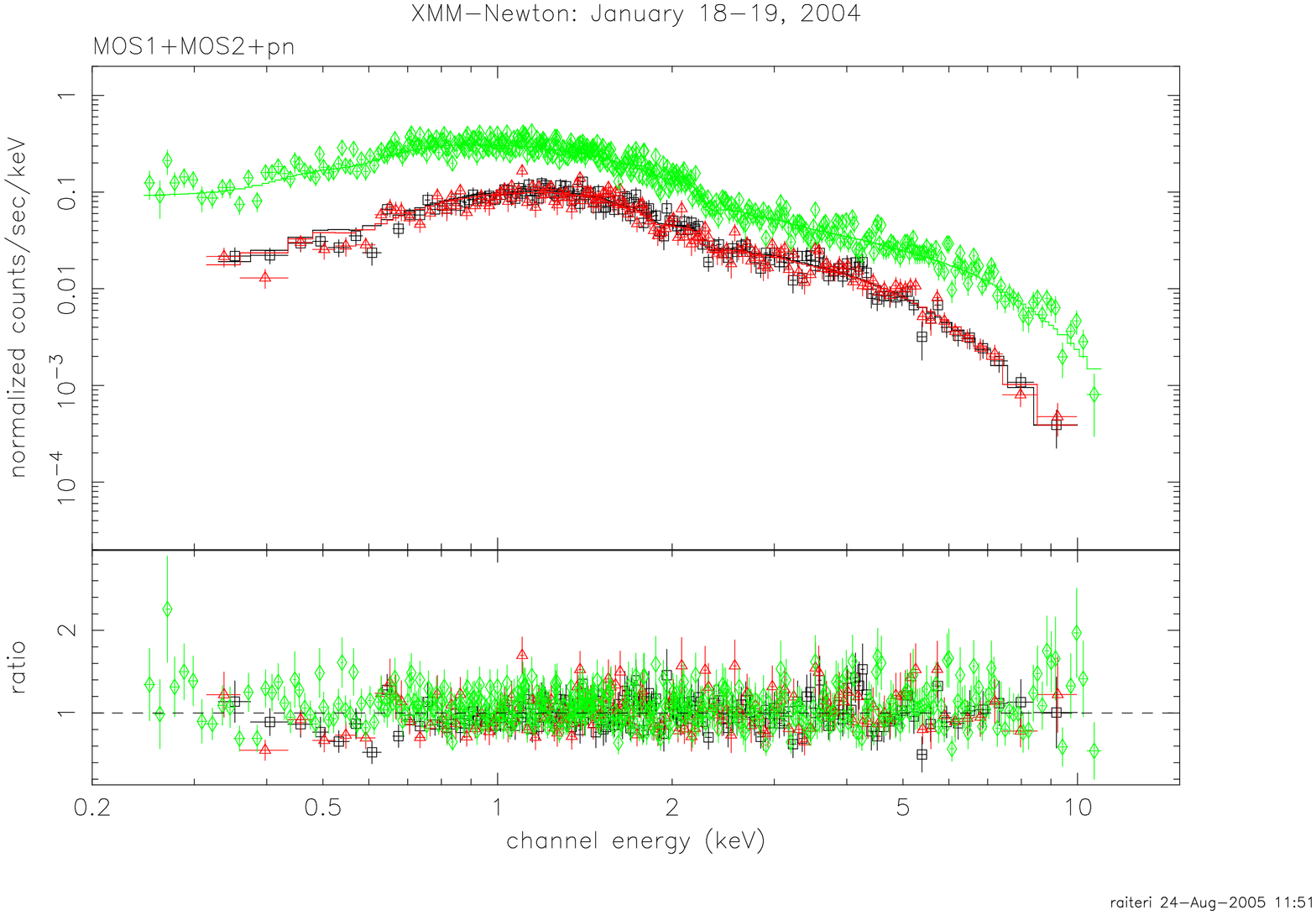}}
   \caption{The X-ray spectrum of AO 0235+16 detected by XMM-Newton on January 18--19, 2004
   fitted by a double power law model with fixed Galactic and $z=0.524$ DLA system absorptions;
   the lower panel shows the ratio of the data to the model fit}
   \label{XMM1}
   \end{figure}

\begin{figure}
   \resizebox{\hsize}{!}{\includegraphics[angle=-90]{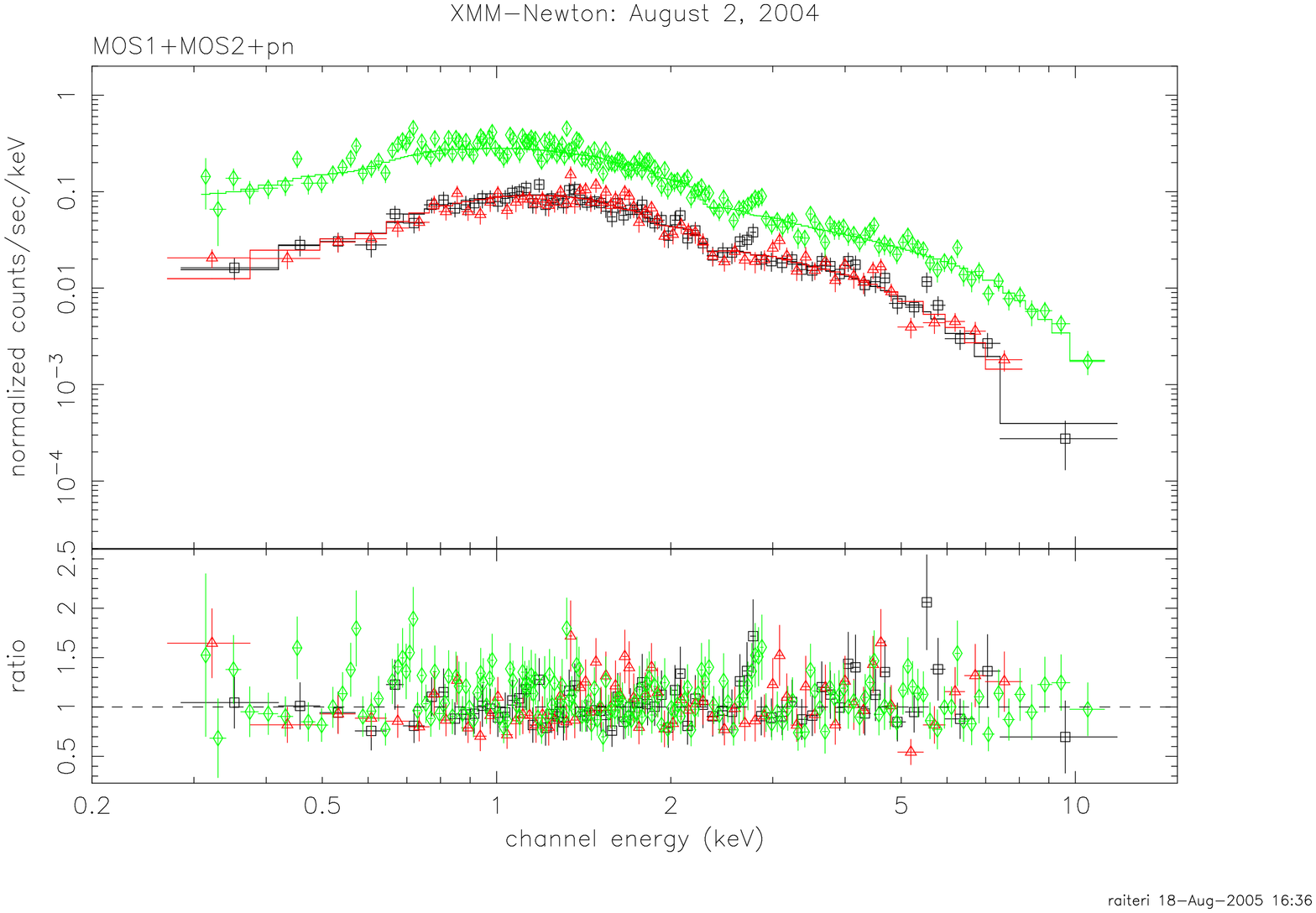}}
   \caption{The X-ray spectrum of AO 0235+16 detected by XMM-Newton on August 2, 2004
   fitted by a double power law model with fixed Galactic and $z=0.524$ DLA system absorptions;
   the lower panel shows the ratio of the data to the model fit}
   \label{XMM2}
   \end{figure}

\begin{figure}
   \resizebox{\hsize}{!}{\includegraphics[angle=-90]{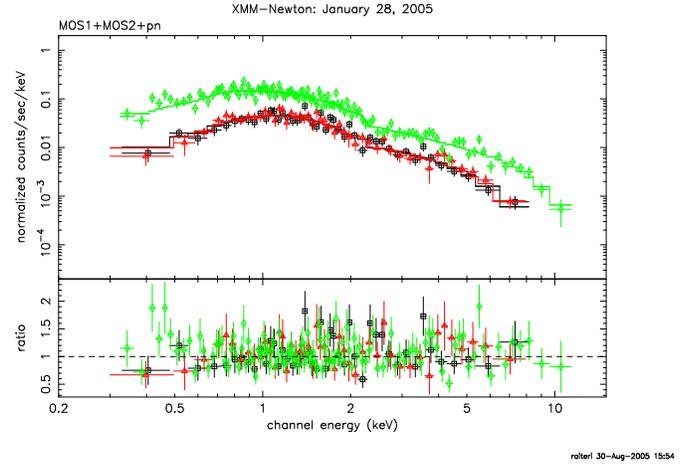}}
   \caption{The X-ray spectrum of AO 0235+16 detected by XMM-Newton on January 28, 2005
   fitted by a double power law model with fixed Galactic and $z=0.524$ DLA system absorptions;
   the lower panel shows the ratio of the data to the model fit}
   \label{XMM3}
   \end{figure}

The resulting spectra are shown in Figs.\ \ref{Chandra}--\ref{XMM3}, fitted by a double power law
$k_1 E^{- \Gamma_1} + k_2 E^{- \Gamma_2}$ 
with Galactic absorption $N_{\rm H}=0.679 \times 10^{21} \, \rm cm^{-2}$
plus extra-absorption by the intervening system at $z=0.524$
$N_{\rm H}=5.0 \times 10^{21} \, \rm cm^{-2}$ (see next section).
The ratio of the data to the model fit is shown in the bottom panels.

In Figs.\ \ref{XMM0}--\ref{XMM3} the MOS1 and MOS2 spectra are plotted with black squares and red triangles, 
respectively, while the pn one is shown with green diamonds.

\subsection{Results}

We start by considering 
\begin{itemize}
\item Model 1: a single power law  $k E^{- \Gamma}$
\end{itemize}
with fixed Galactic photo-electric absorption,
and extra-absorption attributed to the intervening DLA system at $z=0.524$.
We adopted the value $N_{\rm H}=0.679 \times 10^{21} \, \rm cm^{-2}$ for the Galactic hydrogen column, from
the Leiden-Argentine-Bonn Survey{\footnote {\tt http://www.astro.uni-bonn.de/$\sim$webrai/german/tools \_labsurvey.php}}
(Kalberla et al.\ \cite{kal05}),
which is the most sensitive Milky Way \ion{H}{i} survey to date, with the most extensive
coverage both spatially and kinematically.
Both the Galactic and the $z=0.524$ absorber are first assumed to have solar abundance ratios,
as was done in the past data reductions which led to the results summarized in Table \ref{past}.

Table \ref{pow} shows the results for the Chandra and XMM-Newton pointings analysed in the present work (first line).
The fit is good for all cases, but the best-fit $N_{\rm H}$ value of the $z=0.524$ system is nearly half the
$N_{\rm \ion{H}{i}}$ value derived by fitting the Ly$\alpha$ absorption line in the HST spectrum of February 1998, which is
$4.5 \pm 0.4 \times 10^{21} \, \rm cm^{-2}$ according to Turnshek et al.\ (\cite{tur03}), and
$5 \pm 1 \times 10^{21} \, \rm cm^{-2}$ according to Junkkarinen et al.\ (\cite{jun04}).
Notice that the $N_{\rm \ion{H}{i}}$ value derived from the Ly$\alpha$ absorption line accounts for
neutral hydrogen absorption only, and it is thus a lower limit to the $N_{\rm H}$ value derived with {\tt Xspec},
which instead takes into account all the possible absorptions of all elements in molecular, neutral, or ionized form.
Moreover, the Galactic gas is metal-poor with respect to solar,
and a metal-poor chemical composition appears more suitable also to describe the $z=0.524$ absorber.
We thus re-analyse our datasets by using the Wilms et al.\ (\cite{wil00}) model for both
the interstellar medium and the $z=0.524$ absorber.
This model includes, besides revised abundances for the interstellar medium,
updated photoionization cross-sections,
as well as a treatment of interstellar grains and $\rm H_2$ molecules.
For each observation in Table \ref{pow}, 
the results of the corresponding analysis are shown in the second row.
As one can see, there are little differences in the spectral indices and fluxes
with respect to the previous case, but the value of $N_{\rm H}$ increases, as expected,
since now we have less absorbing metals. However, it is
still well below the value of $N_\ion{H}{i}$ 
derived by Turnshek et al.\ (\cite{tur03}) and 
Junkkarinen et al.\ (\cite{jun04})\footnote{When we try more complicated models introducing 
some spectral curvature, the $N_{\rm H}$ values increase a bit more,
but the uncertainties of the parameters are large.}.

We then fix the parameter $N_{\rm H}^{0.524}=5.0 \times 10^{21} \, \rm cm^{-2}$.
As a consequence, the single power law model gives a worse fit
(see third rows in Table \ref{pow}),
especially for the XMM-Newton spectra of February 2002.
Thus, we investigate the reliability of other models implying some spectral curvature.
We consider
\begin{itemize}
\item Model 2: a logarithmic parabola
$k E^{(- \Gamma + \beta \log E)}$ 
(Landau et al.\ \cite{lan86}; Giommi et al.\ \cite{gio02}; Perlman et al.\ \cite{per05})
\item Model 3: a broken power law
$k E^{- \Gamma_1}$ for $E < E_{\rm break}$,
$k E_{\rm break}^{\Gamma_2 - \Gamma_1} E^{- \Gamma_2}$ for $E \ge E_{\rm break}$
\item Model 4: a curved model
$k E^{- \Gamma_1} {( 1 + E/E_{\rm break})}^{\Gamma_1 - \Gamma_2}$
(Fossati et al.\ \cite{fos00}; Tavecchio et al.\ \cite{tav01})
\item Model 5: a double power law
$k_1 E^{- \Gamma_1} + k_2 E^{- \Gamma_2}$
\end{itemize}
all of them with the same two absorption components of Model 1,
with fixed $N_{\rm H}$ values and with the Wilms et al.\ (\cite{wil00})
model for the photo-electric absorption.
The results of these fits are shown in Tables \ref{parabola}--\ref{double}.
The last column reports the F-test probability of the model with respect to the Model 1 case with
$N_{\rm H}^{0.524}=5.0 \times 10^{21} \, \rm cm^{-2}$ (third rows in Table \ref{pow}).
The lower the probability, the better the new model is with respect to the previous one.
As one can see from the statistics, all these models fit the data better than a single power law.
However, there is not a unique, best-fit model for all the available data.

We notice that this spectral curvature is expected from the superposition of the 
high-energy tail of the synchrotron (or thermal, see Sect.\ 6) component on the 
(higher-energy) inverse-Compton branch, both components possibly 
implying power-law spectral shapes in this limited energy range.
Moreover, in the specific case of AO 0235+16 the curvature should be more dramatic,
since previous works have highlighted the existence of a UV--soft-X-ray bump in the SED,
indicating that the soft X-ray spectrum must turn up very abruptly towards the UV spectral region
(Junkkarinen et al.\ \cite{jun04}; Raiteri et al.\ \cite{rai05}).
From this point of view, Model 5, being the sum of two power laws,
appears to be the most suitable model to describe the source spectrum.

\begin{figure}
   \resizebox{\hsize}{!}{\includegraphics{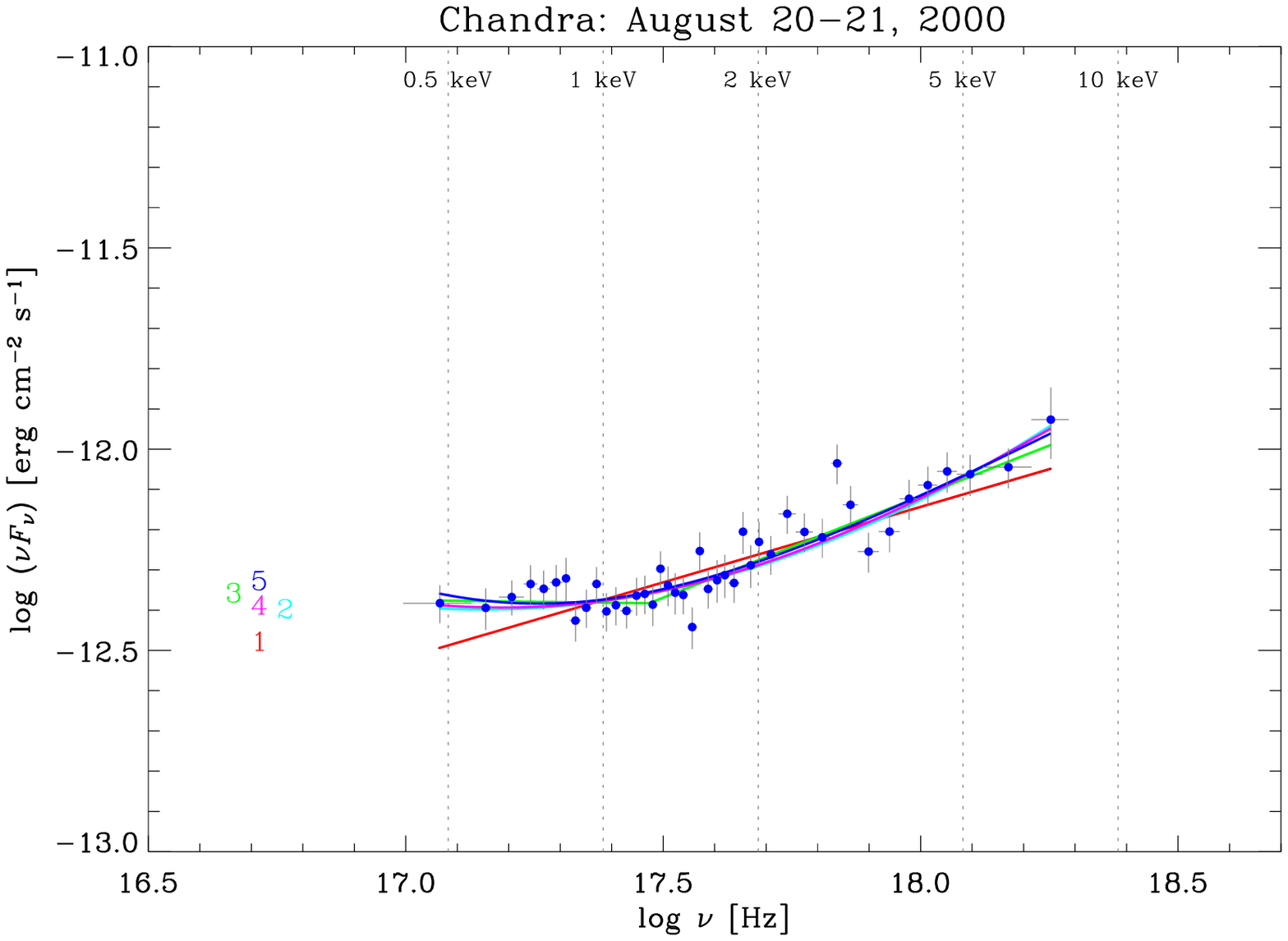}}
   \caption{Fits to the X-ray spectrum of AO 0235+16 during the Chandra observation of August 20--21, 2000
   according to Models 1--5;
   the blue dots represent the unfolded and unabsorbed data obtained by applying Model 5, after some binning}
   \label{sed_Chandra}
   \end{figure}

\begin{figure}
   \resizebox{\hsize}{!}{\includegraphics{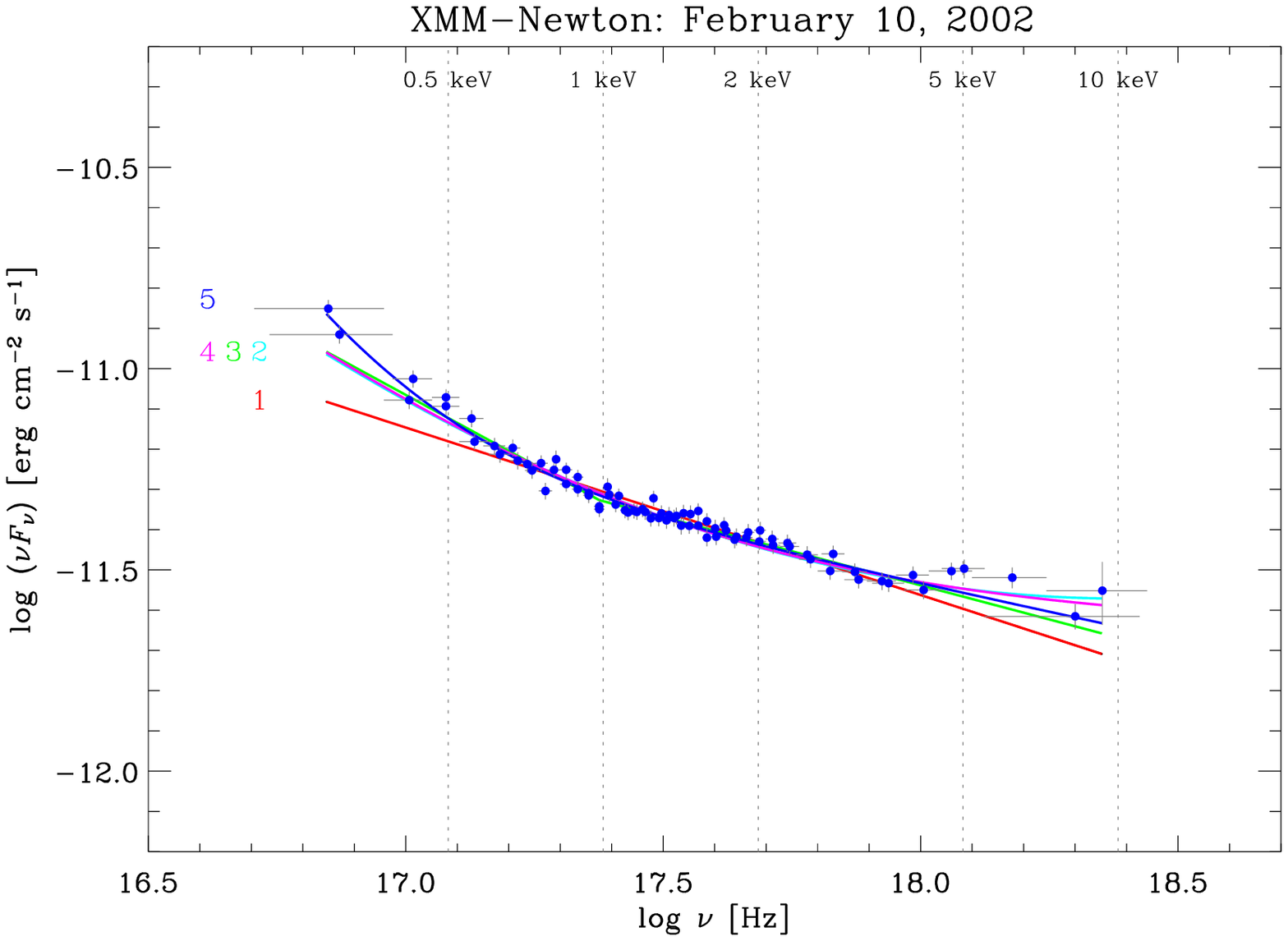}}
   \caption{Fits to the X-ray spectrum of AO 0235+16 during the XMM-Newton observation of February 10, 2002
   according to Models 1--5;
   the blue dots represent the unfolded and unabsorbed data obtained by applying Model 5, after some binning}
   \label{sed_XMM0}
   \end{figure}

\begin{figure}
   \resizebox{\hsize}{!}{\includegraphics{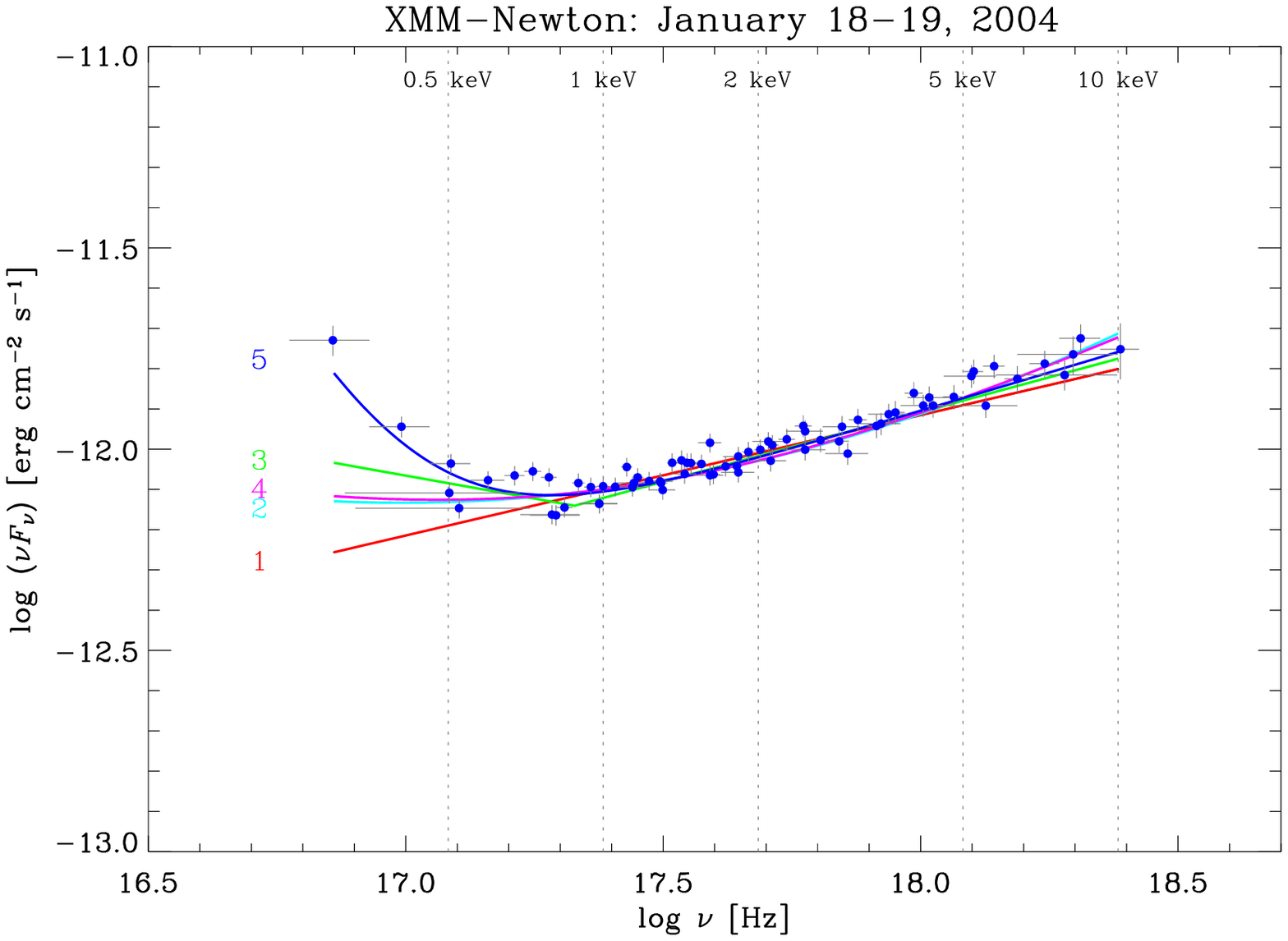}}
   \caption{Fits to the X-ray spectrum of AO 0235+16 during the XMM-Newton observation of January 18--19, 2004
   according to Models 1--5;
   the blue dots represent the unfolded and unabsorbed data obtained by applying Model 5, after some binning}
   \label{sed_XMM1}
   \end{figure}

\begin{figure}
   \resizebox{\hsize}{!}{\includegraphics{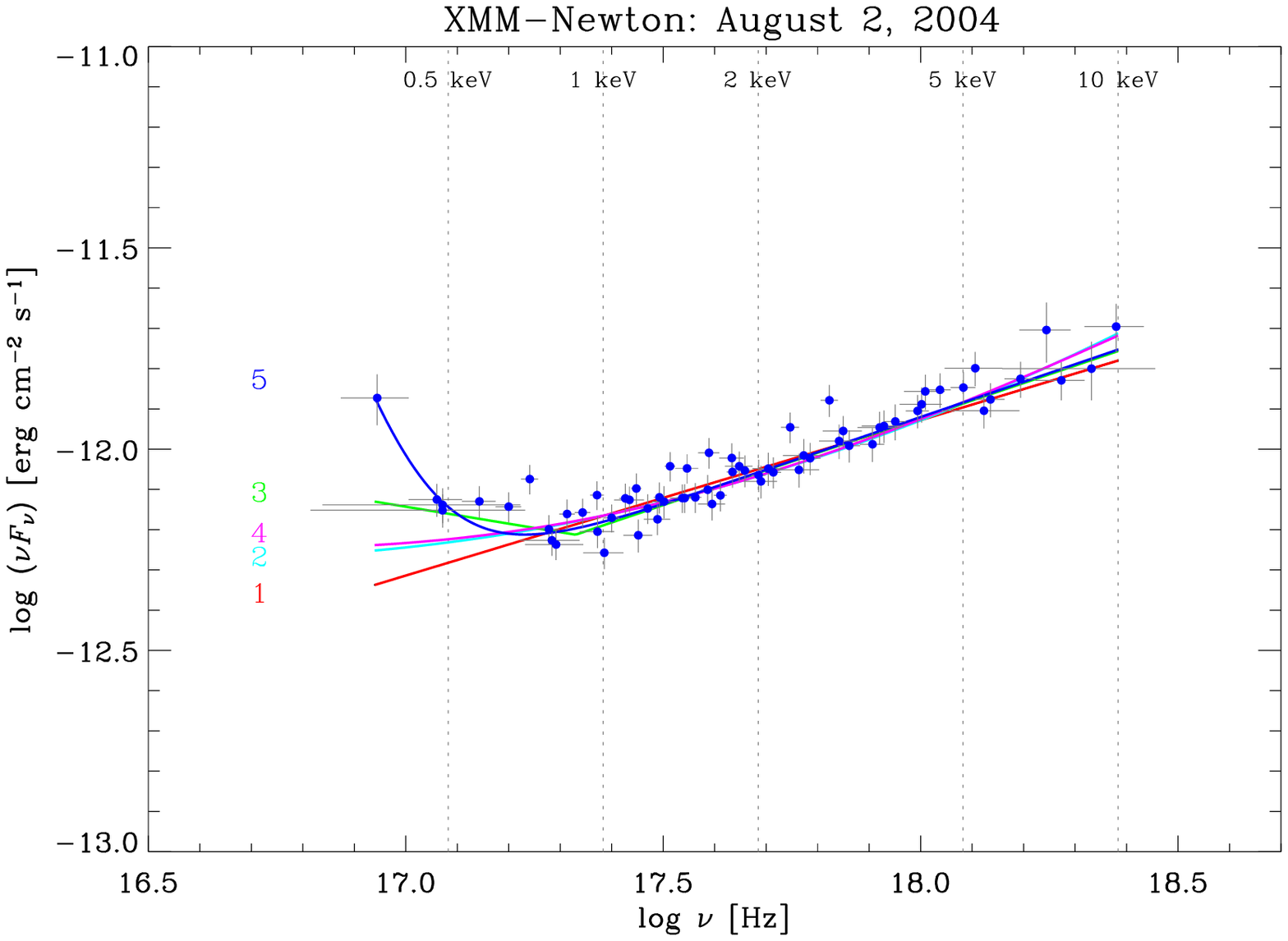}}
   \caption{Fits to the X-ray spectrum of AO 0235+16 during the XMM-Newton observation of August 2, 2004
   according to Models 1--5;
   the blue dots represent the unfolded and unabsorbed data obtained by applying Model 5, after some binning}
   \label{sed_XMM2}
   \end{figure}

\begin{figure}
   \resizebox{\hsize}{!}{\includegraphics{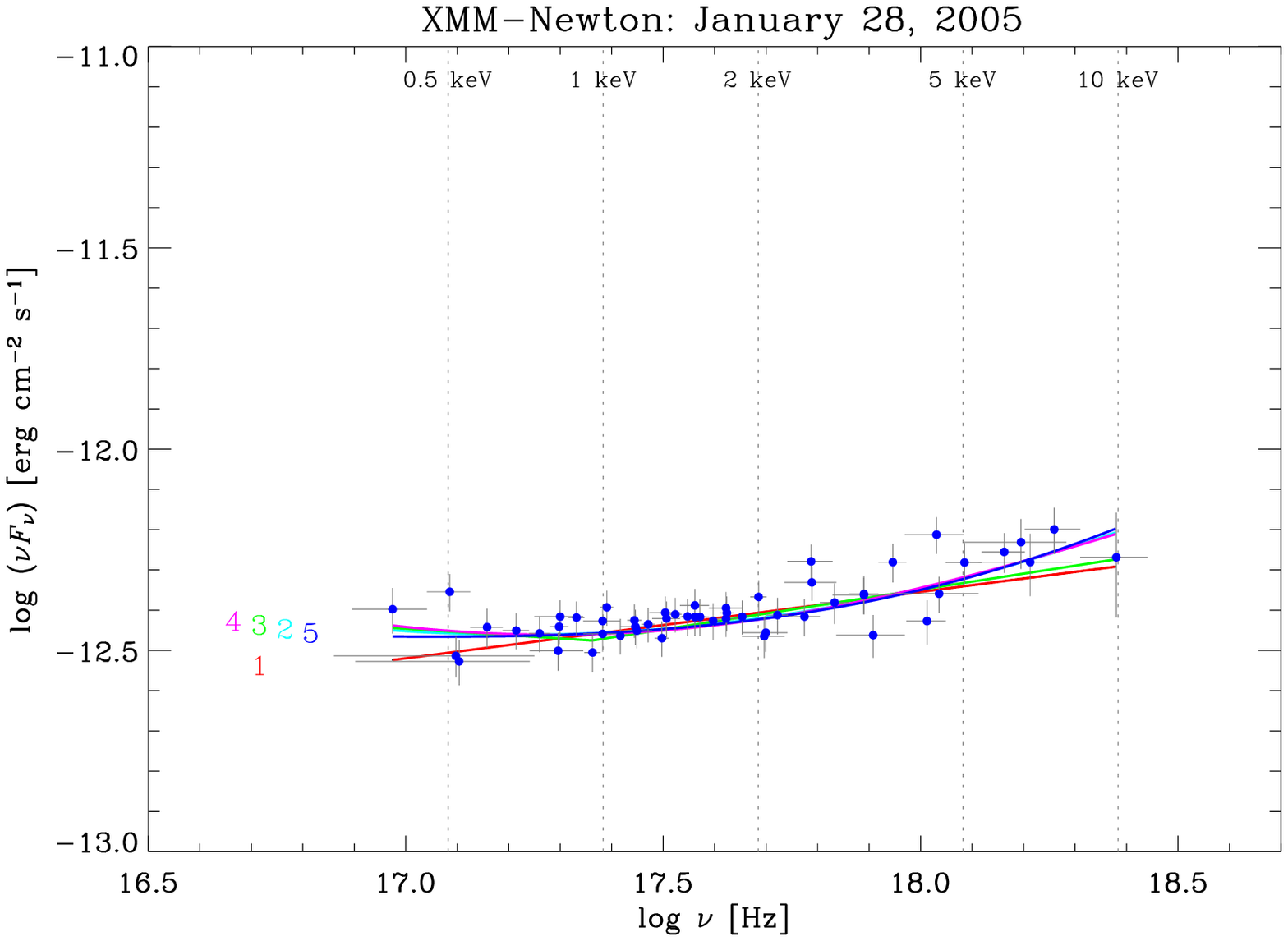}}
   \caption{Fits to the X-ray spectrum of AO 0235+16 during the XMM-Newton observation of January 28, 2005
   according to Models 1--5;
   the blue dots represent the unfolded and unabsorbed data obtained by applying Model 5, after some binning}
   \label{sed_XMM3}
   \end{figure}

The results of our study are also presented in Figs.\ \ref{sed_Chandra}--\ref{sed_XMM3}, where the
model spectra are shown for each observation.
Different model fits are plotted in different colours:
Model 1 (single power law) is represented in red, Model 2 (logarithmic parabola)
in cyan, Model 3 (broken power law) in green,
Model 4 (curved) in pink, and Model 5 (double power law) in blue.
Unfolded and unabsorbed data obtained by applying the double power law model are also plotted, 
after some binning;
we have to warn that had we chosen another model to unfold the data, 
the points would have shown accordingly.

\begin{table*}
\centering
\caption{Results of the X-ray data analysis with
Model 1: a single power law with Galactic absorption of $0.679 \times 10^{21} \, \rm cm^{-2}$
and extra-absorption at $z=0.524$; for each observation the first row is obtained by adopting solar abundances,
while the second row refers to the metal-poor model by Wilms et al.\ (\cite{wil00}), the third row fixing the
extra-absorption at $5.0 \times 10^{21} \, \rm cm^{-2}$. $F_{1 \, \rm keV}$ is the unabsorbed 1 keV flux density;
$F$ is the observed 2--10 keV flux.}
\begin{tabular}{l c c c c c c}
\hline
Satellite & Observation        & $N_{\rm H}^{0.524}$         & $\Gamma$  & $F_{1 \, \rm keV}$  & $F$ (2--10 keV)  & $\chi ^2 / \nu$ $(\nu)$\\
          & date                 &($10^{21} \, \rm cm^{-2}$) &            & ($\mu \rm Jy$)&($10^{-12}$ erg cm$^{-2}$ s$^{-1}$)&  \\
\hline
Chandra       &Aug.\ 2000  & $2.36^{+0.48}_{-0.45}$      & $1.53 \pm 0.07$ & $0.16 \pm 0.01$ & 1.25    & 0.83 (108)\\
              &                & $3.44^{+0.65}_{-0.61}$      & $1.51 \pm 0.07$ & $0.15 \pm 0.01$ & 1.28    & 0.80 (108)\\
              &                & 5.0, fixed                  & $1.62 \pm 0.05$ & $0.17 \pm 0.01$ & 1.19    & 0.92 (109)\\
\hline
XMM           &Feb.\ 2002  & $2.94 \pm 0.12$             & $2.36 \pm 0.03$ & $1.97 \pm 0.04$    & 4.47    & 1.08 (483)\\
              &                & $4.09 \pm 0.17$             & $2.32 \pm 0.02$ & $1.88 \pm 0.04$    & 4.54    & 1.06 (483)\\
              &                & 5.0, fixed                  & $2.42 \pm 0.02$ & $2.04 \pm 0.02$    & 4.26    & 1.21 (484)\\
\hline
XMM           &Jan.\ 2004  & $2.57 \pm 0.20$             & $1.64 \pm 0.03$ & $0.31 \pm 0.01$ & 2.06    & 1.02 (704)\\
              &                & $3.75 \pm 0.28$             & $1.62 \pm 0.03$ & $0.30 \pm 0.01$ & 2.07    & 1.02 (704)\\
              &                & 5.0, fixed                  & $1.70 \pm 0.02$ & $0.33 \pm 0.01$ & 1.99    & 1.08 (705)\\
\hline
XMM           &Aug.\ 2004  & $2.45^{+0.35}_{-0.33}$      & $1.55 \pm 0.04$ & $0.26 \pm 0.01$ & 2.03    & 0.92 (320)\\
              &                & $3.63^{+0.49}_{-0.46}$      & $1.53 \pm 0.04$ & $0.26 \pm 0.01$ & 2.04    & 0.93 (320)\\
              &                & 5.0, fixed                  & $1.61 \pm 0.03$ & $0.28 \pm 0.01$ & 1.96    & 0.98 (321)\\
\hline
XMM           &Jan.\ 2005  & $2.91^{+0.50}_{-0.46}$      & $1.80 \pm 0.07$ & $0.14 \pm 0.01$ & 0.73   & 1.05 (208)\\
              &                & $4.24^{+0.65}_{-0.60}$      & $1.78 \pm 0.06$ & $0.14 \pm 0.01$ & 0.73   & 1.05 (208)\\
              &                & 5.0, fixed                  & $1.83 \pm 0.05$ & $0.14 \pm 0.01$ & 0.71   & 1.06 (209)\\
\hline
\label{pow}
\end{tabular}
\end{table*}

\begin{table*}
\centering
\caption{Results of the X-ray data analysis with
Model 2: a logarithmic parabola with absorption according to the metal-poor model by Wilms et al.\ (\cite{wil00}).
The Galactic hydrogen column is $0.679 \times 10^{21} \, \rm cm^{-2}$, while that of the $z=0.524$ absorber is
$5.0 \times 10^{21} \, \rm cm^{-2}$. $F_{1 \, \rm keV}$ is the unabsorbed 1 keV flux density;
$F$ is the observed 2--10 keV flux.}
\begin{tabular}{l c c c c c c c c}
\hline
Satellite & Observation   & $N_{\rm H}^{0.524}$          & $\Gamma$  & $\beta$ &$F_{1 \, \rm keV}$    & $F$ (2--10 keV)  & $\chi ^2 / \nu$ $(\nu)$ &F-test\\
          & date          &($10^{21} \, \rm cm^{-2}$) &           &         &($\mu \rm Jy$)&($10^{-12}$ erg cm$^{-2}$ s$^{-1}$)&  &prob.\\
\hline
Chandra   &Aug.\ 2000 & 5.0, fixed  & 1.82 & 0.38 & 0.17 & 1.36 & 0.79 (108) & $3.4 \times 10^{-5}$\\
XMM       &Feb.\ 2002 & 5.0, fixed  & 2.51 & 0.25 & 2.01 & 4.71 & 1.07 (483) & $4.6 \times 10^{-15}$\\
XMM       &Jan.\ 2004 & 5.0, fixed  & 1.83 & 0.22 & 0.33 & 2.10 & 1.03 (704) & $5.2 \times 10^{-9}$\\
XMM       &Aug.\ 2004 & 5.0, fixed  & 1.73 & 0.18 & 0.28 & 2.04 & 0.96 (320) & $1.8 \times 10^{-3}$\\
XMM       &Jan.\ 2005 & 5.0, fixed  & 1.94 & 0.19 & 0.14 & 0.75 & 1.04 (208) & $2.3 \times 10^{-2}$\\
\hline
\label{parabola}
\end{tabular}
\end{table*}

\begin{table*}
\centering
\caption{Results of the X-ray data analysis with
Model 3: a broken power law with absorption according to the metal-poor model by Wilms et al.\ (\cite{wil00}).
The Galactic hydrogen column is $0.679 \times 10^{21} \, \rm cm^{-2}$, while that of the $z=0.524$ absorber is
$5.0 \times 10^{21} \, \rm cm^{-2}$. $F_{1 \, \rm keV}$ is the unabsorbed 1 keV flux density;
$F$ is the observed 2--10 keV flux.}
\begin{tabular}{l c c c c c c c c c}
\hline
Satellite & Observation   & $N_{\rm H}^{0.524}$          & $\Gamma_1$  & $\Gamma_2$ & $E_{\rm break}$&$F_{1 \, \rm keV}$ & $F$ (2--10 keV)  & $\chi ^2 / \nu$ $(\nu)$ & F-test\\
          & date          &($10^{21} \, \rm cm^{-2}$) &             &            & (keV)      &($\mu \rm Jy$)&($10^{-12}$ erg cm$^{-2}$ s$^{-1}$)& &prob.\\
\hline
Chandra   &Aug.\ 2000 & 5.0, fixed  & 2.02       & 1.50      & 1.23      & 0.17     & 1.30       & 0.76 (107) & $1.3 \times 10^{-5}$\\
XMM       &Feb.\ 2002 & 5.0, fixed  & 2.70       & 2.34      & 0.98      & 1.94     & 4.51       & 1.06 (482) & $4.3 \times 10^{-15}$\\
XMM       &Jan.\ 2004 & 5.0, fixed  & 2.23       & 1.65      & 0.88      & 0.31     & 2.04       & 1.02 (703) & $8.0 \times 10^{-10}$\\
XMM       &Aug.\ 2004 & 5.0, fixed  & 2.21       & 1.57      & 0.88      & 0.27     & 2.00       & 0.93 (319) & $3.9 \times 10^{-5}$\\
XMM       &Jan.\ 2005 & 5.0, fixed  & 2.08       & 1.80      & 0.95      & 0.14     & 0.72       & 1.06 (207) & 0.223 \\
\hline
\label{broken}
\end{tabular}
\end{table*}

\begin{table*}
\centering
\caption{Results of the X-ray data analysis with
Model 4: a curved model with absorption according to the metal-poor model by Wilms et al.\ (\cite{wil00}).
The Galactic hydrogen column is $0.679 \times 10^{21} \, \rm cm^{-2}$, while that of the $z=0.524$ absorber is
$5.0 \times 10^{21} \, \rm cm^{-2}$. $F_{1 \, \rm keV}$ is the unabsorbed 1 keV flux density;
$F$ is the observed 2--10 keV flux.}
\begin{tabular}{l c c c c c c c c c}
\hline
Satellite & Observation   & $N_{\rm H}^{0.524}$       & $\Gamma_1$  & $\Gamma_2$ & $E_{\rm break}$&$F_{1 \, \rm keV}$ & $F$ (2--10 keV)  & $\chi ^2 / \nu$ $(\nu)$ & F-test\\
          & date          &($10^{21} \, \rm cm^{-2}$) &             &            & (keV)      &($\mu \rm Jy$)&($10^{-12}$ erg cm$^{-2}$ s$^{-1}$)& &prob.\\
\hline
Chandra   &Aug.\ 2000 & 5.0, fixed & 2.59 & 1.09 & 0.99 & 0.17 & 1.35 & 0.79 (107) & $1.2 \times 10^{-4}$\\
XMM       &Feb.\ 2002 & 5.0, fixed & 3.00 & 2.02 & 1.00 & 2.01 & 4.68 & 1.07 (482) & $3.3 \times 10^{-14}$\\
XMM       &Jan.\ 2004 & 5.0, fixed & 2.30 & 1.39 & 0.98 & 0.33 & 2.10 & 1.03 (703) & $1.6 \times 10^{-8}$\\
XMM       &Aug.\ 2004 & 5.0, fixed & 2.14 & 1.36 & 0.95 & 0.28 & 2.04 & 0.96 (319) & $4.1 \times 10^{-3}$\\
XMM       &Jan.\ 2005 & 5.0, fixed & 2.38 & 1.53 & 1.00 & 0.14 & 0.75 & 1.05 (207) & $8.2 \times 10^{-2}$\\
\hline
\label{curved}
\end{tabular}
\end{table*}

\begin{table*}
\centering
\caption{Results of the X-ray data analysis with
Model 5: a double power law with absorption according to the metal-poor model by Wilms et al.\ (\cite{wil00}).
The Galactic hydrogen column is $0.679 \times 10^{21} \, \rm cm^{-2}$, while that of the $z=0.524$ absorber is
$5.0 \times 10^{21} \, \rm cm^{-2}$. $F_{1 \, \rm keV}$ is the unabsorbed 1 keV flux density;
$F$ is the observed 2--10 keV flux.}
\begin{tabular}{l c c c c c c c c}
\hline
Satellite & Observation   & $N_{\rm H}^{0.524}$       & $\Gamma_1$  & $\Gamma_2$ &$F_{1 \, \rm keV}$ & $F$ (2--10 keV)  & $\chi ^2 / \nu$ $(\nu)$ & F-test\\
          & Date          &($10^{21} \, \rm cm^{-2}$) &             &            &($\mu \rm Jy$)&($10^{-12}$ erg cm$^{-2}$ s$^{-1}$)& &prob.\\
\hline
Chandra   &Aug.\ 2000 & 5.0, fixed & 2.86 & 1.33 & 0.17 & 1.34 & 0.79 (107) & $1.4 \times 10^{-4}$\\
XMM       &Feb.\ 2002 & 5.0, fixed & 4.18 & 2.27 & 1.99 & 4.58 & 1.06 (482) & $2.9 \times 10^{-15}$\\
XMM       &Jan.\ 2004 & 5.0, fixed & 4.39 & 1.62 & 0.32 & 2.07 & 1.02 (703) & $4.2 \times 10^{-9}$\\
XMM       &Aug.\ 2004 & 5.0, fixed & 6.11 & 1.56 & 0.27 & 2.02 & 0.92 (319) & $1.6 \times 10^{-5}$\\
XMM       &Jan.\ 2005 & 5.0, fixed & 1.14 & 2.08 & 0.14 & 0.75 & 1.04 (207) & $6.9 \times 10^{-2}$ \\
\hline
\label{double}
\end{tabular}
\end{table*}

\subsection{Comparison with other works}

By comparing the results of Table \ref{pow} with those of Table \ref{past}
we notice that the two ASCA pointings of February 1994 and 1998 detected the source at flux levels
comparable to those observed by XMM-Newton in January and August 2004 
($F_{1 \, \rm keV} \sim 0.3 \mu \rm Jy$). The ASCA data were analyzed by Madejski et al.\ (\cite{mad96})
and Junkkarinen et al.\ (\cite{jun04})
with prescriptions for absorption similar to ours, hence the results can easily be compared,
highlighting how the source can exhibit noticeable spectral variations 
(from $\Gamma \sim 1.6$ to $\Gamma \sim 2.0$) at the same brightness level.

A previous analysis of the August 2000 Chandra observation was published by Turnshek et al.\ (\cite{tur03}).
The interest of these authors was to study the metallicity of the intervening DLA system at $z=0.524$.
They fit the Chandra spectrum with a power law in the 2--8 keV energy range,
where absorption is expected not to affect the continuum flux,
in order to fix both the slope and normalization of the X-ray spectrum.
A best-fit photon index $\Gamma = 1.61 \pm 0.10$ was found.
They adopted a Galactic \ion{H}{i} column density
of $0.87 \times 10^{21} \, \rm cm^{-2}$ after Hartmann \& Burton (\cite{har97}) 
and a value of $N_{\ion{H}{i}}=4.5 \pm 0.4 \times 10^{21} \, \rm cm^{-2}$ for the $z=0.524$ absorber, 
which was derived from a fit of the Ly$\alpha$ absorption line in the February 1998 HST spectrum.
Then they tried different prescriptions for the metal content of both the Galaxy and DLA system,
finding a range of possible metallicity for this latter of 10--40\% of the solar one.
The results of Turnshek et al.\ (\cite{tur03}) are in fair agreement with those we found
by applying a similar model
(single power law with $N_{\rm H}^{0.524} = 5.0 \times 10^{21} \, \rm cm^{-2}$, see Table \ref{pow}).

Foschini et al.\ (\cite{fos06}) analysed the data of the XMM-Newton observation of February 2002.
When fitting a single power law with Galactic absorption in the 1--10 keV energy range, 
they obtained $\Gamma = 2.28 \pm 0.02$,
roughly consistent with the results reported in Table \ref{pow}.

\subsection{Iron lines}

The detection of K$\alpha$ iron fluorescent emission lines in AGN X-ray spectra
is a very important and unique tool to study the immediate vicinity of the
supermassive black hole which most likely is the central engine of the AGN
itself (see e.g.\ the review papers by Fabian et al.\ \cite{fab00} and Reynolds \& Nowak \cite{rey03}).
Up to now, K$\alpha$ iron fluorescent lines have been detected in several Seyfert
galaxies (see the above reviews and references therein) and in some radio
galaxies (e.g.\ Eracleous et al.\ \cite{era00}). As far as we are aware, among blazars,
only 3C 273 has shown this line in its X-ray spectrum (e.g.\ Yaqoob \& Serlemitsos
\cite {yaq00}), and it has been never seen in BL Lac objects, except for the possible
case of S5 0716+71 (Kadler et al.\ \cite{kad05a}).
If BL Lac objects have an accretion disc surrounded by a hot corona and if this
disc is cold enough to allow the emission of Fe K$\alpha$ fluorescent line, then
this line could be hidden by the beamed X-ray continuum of the jet (of
synchrotron or inverse-Compton nature), just like the softer (IR-to-UV) emission
lines are swamped by the corresponding jet continuum. Alternatively, the
disc in BL Lacs, if any, could be hot, or optically thin, or highly ionized,
thus not producing the K$\alpha$ feature.

AO 0235+16 has been classified as a BL Lac object. However, it has shown both
narrow and broad emission lines in its optical spectrum (Cohen et al.\ \cite{coh87};
Nilsson et al.\ \cite{nil96}; Raiteri et al.\ \cite{rai06}),
in particular a prominent \ion{Mg}{ii} broad line. These features
are obviously better seen when the optical continuum from the jet is lower.
The four X-ray spectra taken with Chandra in 2000 and with XMM-Newton in 2004--2005
show the source in a very faint X-ray state.
In both the Chandra and the August 2004 XMM-Newton spectra (see Figs.\ \ref{Chandra} and \ref{XMM2})
a possible emission feature is present around 2.8 keV (and a more marginal one around 3.1 keV).
By adding a Gaussian line of 0.1 keV width in the energy interval 2.7--3.2 keV
to Model 1 (case with $N_{\rm H}^{0.524} = 5.0 \times 10^{21} \, \rm cm^{-2}$),
the best fit is reached when the line is at 2.8 keV and implies some improvement of $\chi ^2$:
$\Delta \chi ^2=2.63$ for the Chandra observation, and 2.25 in the case of the August 2004
XMM-Newton one. The corresponding equivalent widths are $\sim 61$ eV and $\sim 35$ eV, respectively,
but with large uncertainties. Consequently, we can only say that a line may be present in these
two observations, with EW less than $\sim 120$ eV and $\sim 75$ eV, respectively.
This possible emission line in the source frame
would have an energy of about 5.4 keV, a bit far from the rest-frame
energy of 6.40 keV of the cold Fe K$\alpha$ fluorescent line.
However, it is well known that, if the iron lines are produced in the inner
accretion disc, they are affected from both Doppler and gravitational relativistic
effects. Indeed, although they are intrinsically narrow, they can be broadened
and skewed by the large orbital velocities of the accreting material, and
redshifted by the strong gravitation field of the black hole.
In fact, iron lines have been detected peaking at energies from 6--7 keV down to
well below 6 keV in the source frame, also with a variety of shapes and widths
(see the review by Fabian et al.\ \cite{fab00} and references therein).
Even in the same source, the detected line can show dramatic changes in flux, profile, and
redshift with time, as in the emblematic case of the Seyfert 1 galaxy MCG
$-6$-30-15 (e.g.\ Iwasawa et al.\ \cite{iwa96}).

\section{Ultraviolet to X-ray spectral energy distribution}

Besides the X-ray detectors, the XMM-Newton satellite also carries an Optical Monitor 
(OM), which allows
to obtain photometric data in the optical $U$, $B$, $V$ and ultraviolet UV$W1$, UV$M2$, UV$W2$ bands
simultaneously with the X-ray observations.
Optical and UV grisms are also available.

The optical and ultraviolet emission from AO 0235+16 is contaminated by the emission of
ELISA, the AGN which is located only about 2 arcsec southward of the source. The ELISA contribution
becomes more important as long as AO 0235+16 gets fainter, and as long as higher frequencies are considered
(see discussion in Raiteri et al.\ \cite{rai05}).
The XMM-Newton source magnitudes listed in the OM pipeline products
are obtained by performing aperture photometry with a
12 pixel (about 5.7 arcsec) aperture radius, so that they include the contribution of ELISA.
In order to measure the magnitude of AO 0235+16 avoiding the southern AGN contamination,
we analysed the OM images by following the same
procedure adopted in Raiteri et al.\ (\cite{rai05}), 
i.e.\ by using the {\tt IRAF} routine {\tt phot} with an aperture radius of 2 pixels.
We performed differential photometry with respect to Stars 1, 2, 3 (Smith et al.\ \cite{smi85}), and 6
(Fiorucci et al.\ \cite{fio98}), using for these stars the standard magnitudes listed
in the pipeline products.
The source magnitude was derived by averaging the four values obtained with respect to the reference stars.
Table \ref{ommag} reports the inferred optical--UV magnitudes of AO 0235+16 during the XMM-Newton pointings.

\begin{table*}
\centering
\caption{Optical Monitor standard magnitudes of AO 0235+16; numbers in brackets are $1 \sigma$ errors.}
\begin{tabular}{l c c c c c}
\hline
Date               & $V$          & $B$          & $U$          & UV$W1$       & UV$M2$\\
\hline
Feb.\ 10, 2002     &              &              &              & 17.78 (0.06) &             \\
Jan.\ 18--19, 2004$^{a}$& 19.22 (0.07) & 20.09 (0.07) & 20.12 (0.10) & 19.95 (0.11) & 19.68 (0.22)\\
Aug.\ 02, 2004      &              &              & 19.60 (0.08) & 19.97 (0.13) & 19.80 (0.27)\\
Jan.\ 28, 2005     &              &              & 20.24 (0.11) & 20.24 (0.13) & 19.90 (0.20)\\
\hline
\multicolumn{6}{l}{$^{a}$ from Raiteri et al.\ (\cite{rai05})}
\label{ommag}
\end{tabular}
\end{table*}

\begin{figure}
   \resizebox{\hsize}{!}{\includegraphics{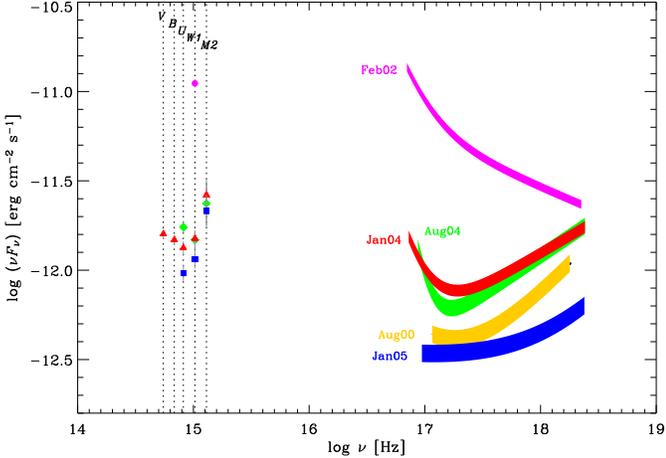}}
   \caption{\ Ultraviolet to X-ray spectral energy distributions of AO 0235+16.
   The OM optical--UV data have been obtained by differential
   photometry with a $\sim 1$ arcsec aperture radius to avoid the contamination by ELISA;
   the pink dot refers to February 10, 2002, red triangles to January 18--19, 2004,
   green diamonds to August 2, 2004, and blue squares to January 28, 2005.
   The X-ray spectra are shown according to Model 5 (double power law),
   the spectrum thickness representing the standard deviation of the data from the model fit,
   and the colours being the same of the corresponding OM data.
   The orange spectrum is from Chandra.}
   \label{sed_con_OM}
   \end{figure}

The ultraviolet to X-ray spectral energy distributions are shown in Fig.\ \ref{sed_con_OM}.
The OM magnitudes of Table \ref{ommag} have been dereddened according to Junkkarinen et al.\
(\cite{jun04}), using the extinction values listed in Table 5, Col.\ 5
of Raiteri et al.\ (\cite{rai05}).
Then the OM fluxes have been obtained with respect to Vega.
Notice that in Fig.\ \ref{sed_con_OM} we have lowered the UV$M2$ flux by 10\% to take into account
the fact that at this frequency the Junkkarinen et al.\ (\cite{jun04}) prescriptions
overestimate the extinction.
   The X-ray spectra are shown according to Model 5 (double power law),
   the spectrum thickness representing the standard deviation of the data from the model fit,
   which is $\sigma=0.049$ for the Chandra pointing of
   August 2000, and $\sigma=0.024, 0.033, 0.046$, and 0.065 for the XMM-Newton pointings of
   February 2002, January 2004, August 2004, and January 2005, respectively.

One can see how the optical--UV state is higher when the X-ray state is higher.
Moreover, the January and August 2004 data show how the steep optical spectrum has
a minimum in the $U$--UV$W1$ frequency range and then the spectrum becomes hard in the UV band,
which is also confirmed by the January 2005 data.
In other words, the OM data indicate that, besides the expected synchrotron peak
likely falling in the infrared band, the SED also presents a UV--soft-X-ray bump.

The existence of a  UV--soft-X-ray bump further justifies the search for a curvature in the
X-ray spectrum, and the preference given to the double power law model.

\section{Interpretation of the UV--soft-X-ray bump}

The  existence of an SED extra-component in the UV--soft-X-ray energy range has already been noticed by
Junkkarinen et al.\ (\cite{jun04}) when analysing the February 1998 ASCA spectrum together with
a contemporaneous optical--UV HST spectrum, and by Raiteri et al.\ (\cite{rai05}), when considering the
January 2004 XMM-Newton data in conjunction with simultaneous optical data acquired at the 2.56 m Nordic Optical Telescope
in the ambit of the WEBT campaign.
Even if Junkkarinen et al.\ (\cite{jun04}) warned that
it could arise from a different behaviour of extinction in the DLA intervening system with respect
to the predicted one, we think that the XMM-Newton UV and X-ray observations indicate
that the existence of a real bump must be considered.

This bump recalls the ``big blue bump" found in the SEDs of many AGNs, and which is usually
ascribed to thermal emission from the accretion disc feeding the supermassive black hole, even if
accretion disc models generally predict a lower flux in the bump tails
(see e.g.\ Krolik \cite{kro99}, and Blaes et al.\ \cite{bla01} for the specific case of the blazar 3C 273).
It is thus interesting to see whether a disc model could explain the bump we see in the SED of AO 0235+16.
By looking at Fig.\ \ref{sed_con_OM} one can infer that this bump would likely peak at 
$\log (\nu F_\nu /\rm erg \, cm^{-2} \, s^{-1} ) > -11.5$, which implies a peak luminosity
$\nu L_\nu = 4 \pi d^2 \nu F_\nu > 10^{46} \, \rm erg \, s^{-1},$ if we assume a flat cosmology with
$H_0=71 \, \rm km \, s^{-1} \, Mpc^{-1}$ and $\Omega=0.27$, leading to a source distance $d \sim 6 \times 10^9 \, \rm pc$.
The peak frequency would likely be in the range $\log (\nu_{\rm peak /Hz}) = 15.5$--16, which in the source rest frame
translates into $\nu'_{\rm peak} \sim (0.6$--$2) \times 10^{16}$ Hz.
According to the non-LTE disc models computed by Hubeny et al.\ (\cite{hub00}) the above 
requirements can be met only in case of a rather massive central black hole ($\ga 10^9 \, \rm M_\odot$)
and a high mass accretion rate ($\ga 2 \, \rm M_\odot \, yr^{-1}$).

The lack of optical--UV spectral information during the XMM-Newton pointing of February 2002 does not allow to
state that the UV--soft-X-ray bump is a strongly variable component.
In any case, a variable bump could be obtained in the accretion disc scenario with changes of the
model parameters, for instance the accretion rate.
A more detailed comparison between the observed SEDs and the accretion disc models is beyond the aim of the present
paper; for the moment we can say that we cannot rule out that the UV--soft-X-ray bump we observe in the AO 0235+16 SEDs
is due to thermal emission from an accretion disc.
The Fe K$\alpha$ emission line tentatively detected in the X-ray spectra would be a signature of this accretion disc.

But other scenarios are also possible.
In a recent paper Ostorero et al.\ (\cite{ost04}) showed how a bump in the UV--soft-X-ray region 
can be obtained in the ambit of a
rotating helical jet model (Villata \& Raiteri \cite{vil99}), by admitting a synchrotron contribution to the soft
X-ray radiation. This contribution would come from the first part of the inner emitting jet,
while the major synchrotron emission at the lower frequencies
comes from just outside. This is a possible picture for the explanation of the UV--soft-X-ray bump
suggested by the XMM-Newton observations.
We could add that, since the ``observed" bump seems to be more pronounced than that predicted by 
Ostorero et al.\ (\cite{ost04}),
in the framework of the helical jet model the two synchrotron components in the SED would come from regions
separated by a sudden variation of opacity.
A similar interpretation was suggested by Raiteri \& Villata (\cite{rai03}) to explain the SED discontinuity
of Mkn 501.

Furthermore, the existence of an additional SED component besides the
classical synchrotron and inverse-Compton ones has recently been postulated for BL Lacertae.
Indeed, when constructing the SED of BL Lac during the BeppoSAX observation
of October--November 2000 with radio data and optical data taken by the WEBT collaboration
(Villata et al.\ \cite{vil02}), Ravasio et al.\ (\cite{rav03}) found a misalignment
of the soft X-ray spectrum with respect to the extrapolation of the optical one.
Even if according to B\"ottcher \& Reimer (\cite{boe04}) this misalignment may be due to the time averaging performed,
it is interesting here to briefly recall the four different hypotheses discussed by 
Ravasio et al.\ (\cite{rav03}) to explain this feature:
1) a higher than Galactic dust-to-gas ratio towards the source (in the BL Lac case there is no intervening system);
2) the detection of bulk Compton emission, i.e.\ radiation produced by inverse-Compton scattering
of UV photons from the disc and the broad-line region off ``cold" electrons;
3) the presence of two synchrotron emitting regions at different distances from the nucleus;
4) the detection of a Klein-Nishina effect on the synchrotron spectrum.

A problem in modelling the absorption towards the source, similarly to
explanation number 1) for BL Lac, cannot be strictly ruled out also for AO 0235+16
(see Junkkarinen et al.\ \cite{jun04}). 
Explanation 2), which requires rather ad-hoc conditions, seems unlikely for AO 0235+16,
for which the very high bulk Lorentz factor would push the inverse-Compton emission at too high energies.
Explanation 3) is similar to the one we favour, while explanation 4), which is
further discussed by Moderski et al.\ (\cite{mod05}), is also a possible alternative interpretation.

\section{X-ray light curves}

As mentioned in the Introduction, AO 0235+16 has shown IDV at various frequencies.
During each of the satellite pointings presented in this paper, we searched for
possible rapid variations of the X-ray flux.

\begin{figure}
   \resizebox{\hsize}{!}{\includegraphics{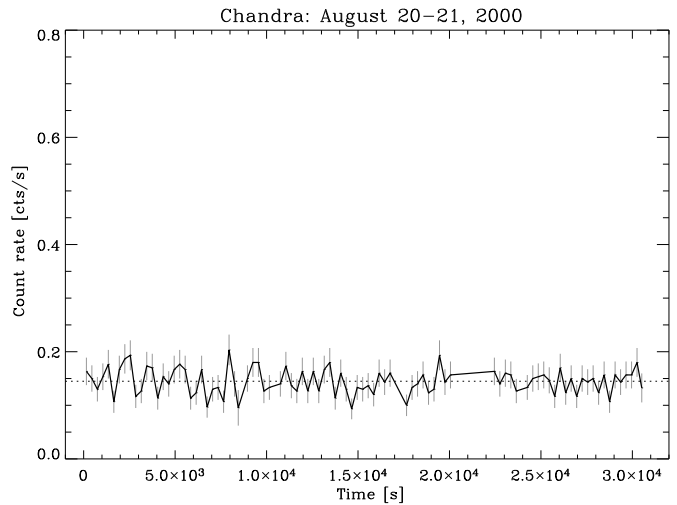}}
   \caption{X-ray light curve extracted from the Chandra observation of August 20--21, 2000;
   data have been binned over 300 s; the horizontal dotted line indicates the average value of the dataset}
   \label{Chandra_lc}
   \end{figure}

\begin{figure}
   \resizebox{\hsize}{!}{\includegraphics{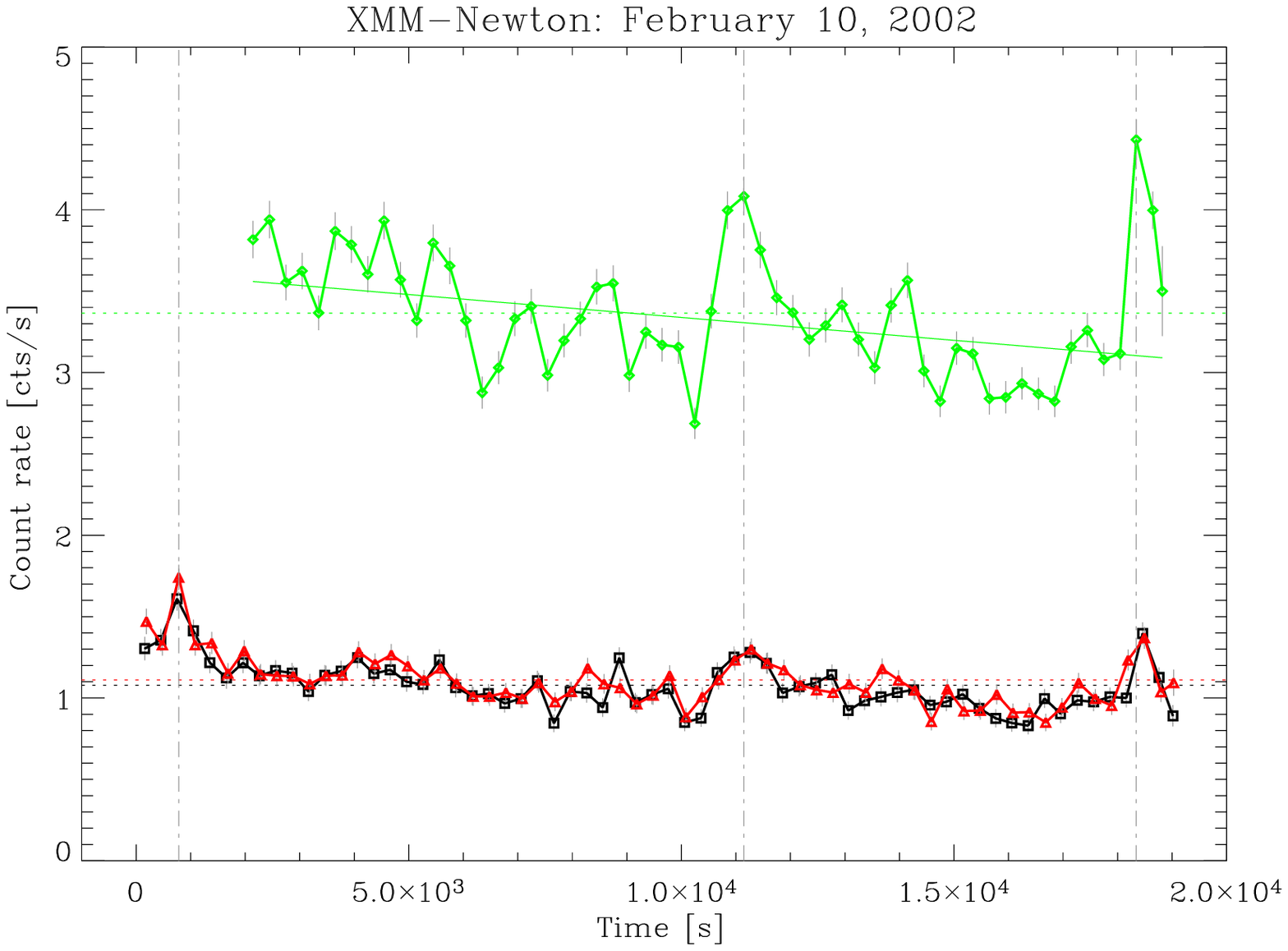}}
   \caption{X-ray light curves extracted from the XMM-Newton observation of February 10, 2002;
   data from pn (green diamonds), MOS1 (black squares), and MOS2 (red triangles) have been binned over 300 s;
   horizontal dotted lines indicate the average values of the datasets; vertical lines highlight the main peaks}
   \label{XMM0_lc}
   \end{figure}

\begin{figure}
   \resizebox{\hsize}{!}{\includegraphics{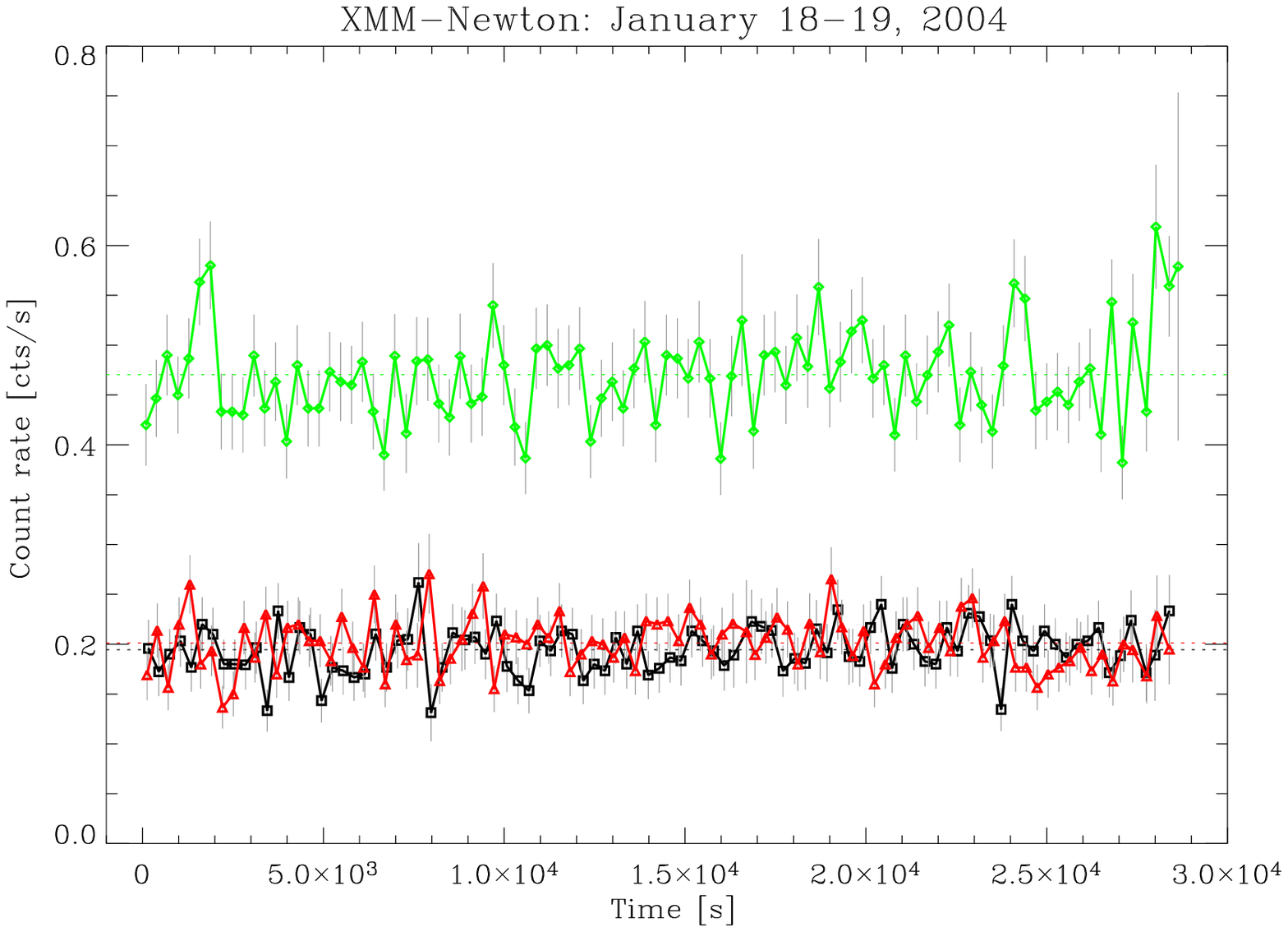}}
   \caption{X-ray light curves extracted from the XMM-Newton observation of January 18--19, 2004;
   data from pn (green diamonds), MOS1 (black squares), and MOS2 (red triangles) have been binned over 300 s;
   horizontal dotted lines indicate the average values of the datasets; vertical lines highlight the main peaks}
   \label{XMM1_lc}
   \end{figure}

\begin{figure}
   \resizebox{\hsize}{!}{\includegraphics{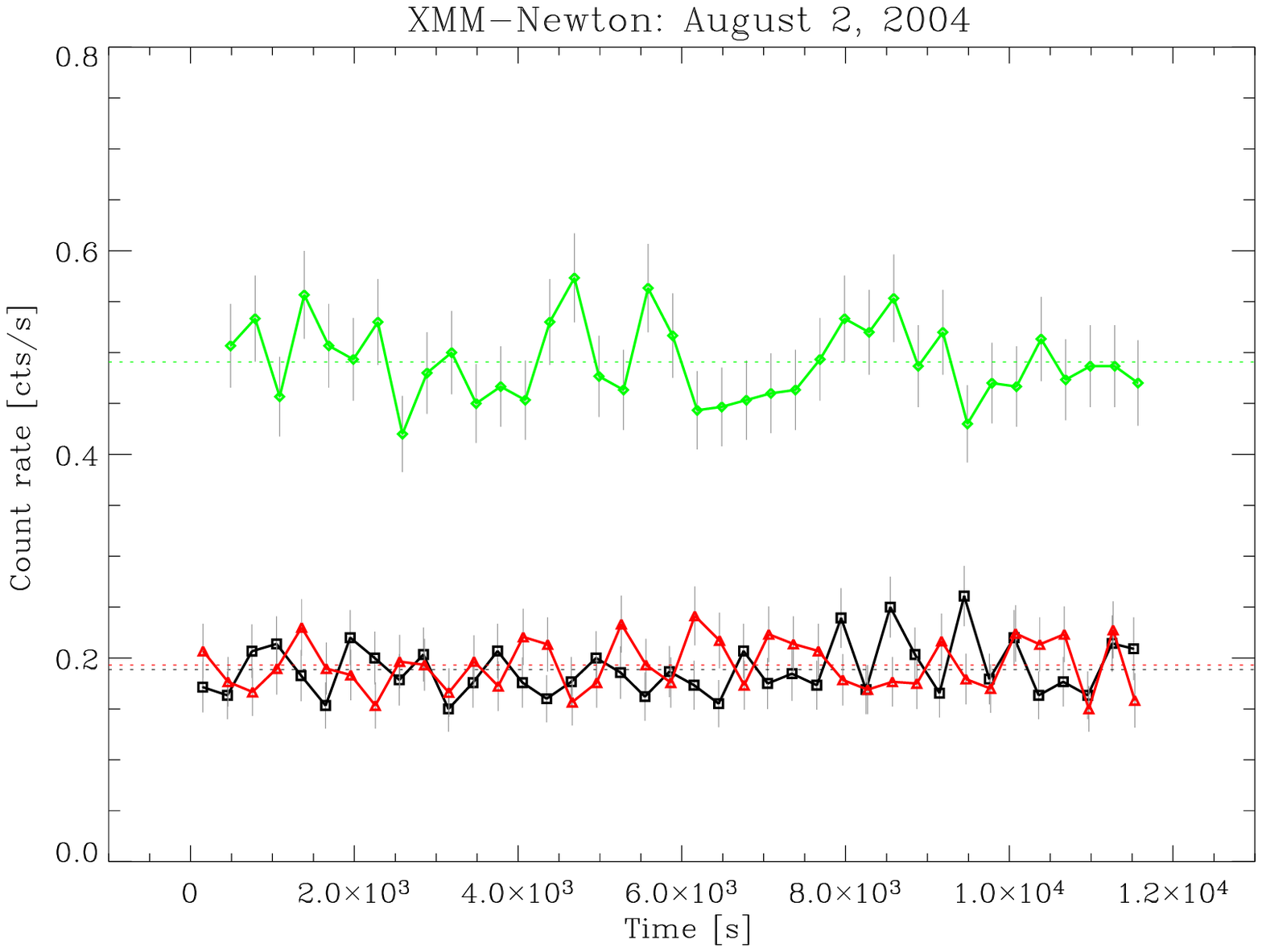}}
   \caption{X-ray light curves extracted from the XMM-Newton observation of August 2, 2004;
   data from pn (green diamonds), MOS1 (black squares), and MOS2 (red triangles) have been binned over 300 s;
   horizontal dotted lines indicate the average values of the datasets; vertical lines highlight the main peaks}
   \label{XMM2_lc}
   \end{figure}

\begin{figure}
   \resizebox{\hsize}{!}{\includegraphics{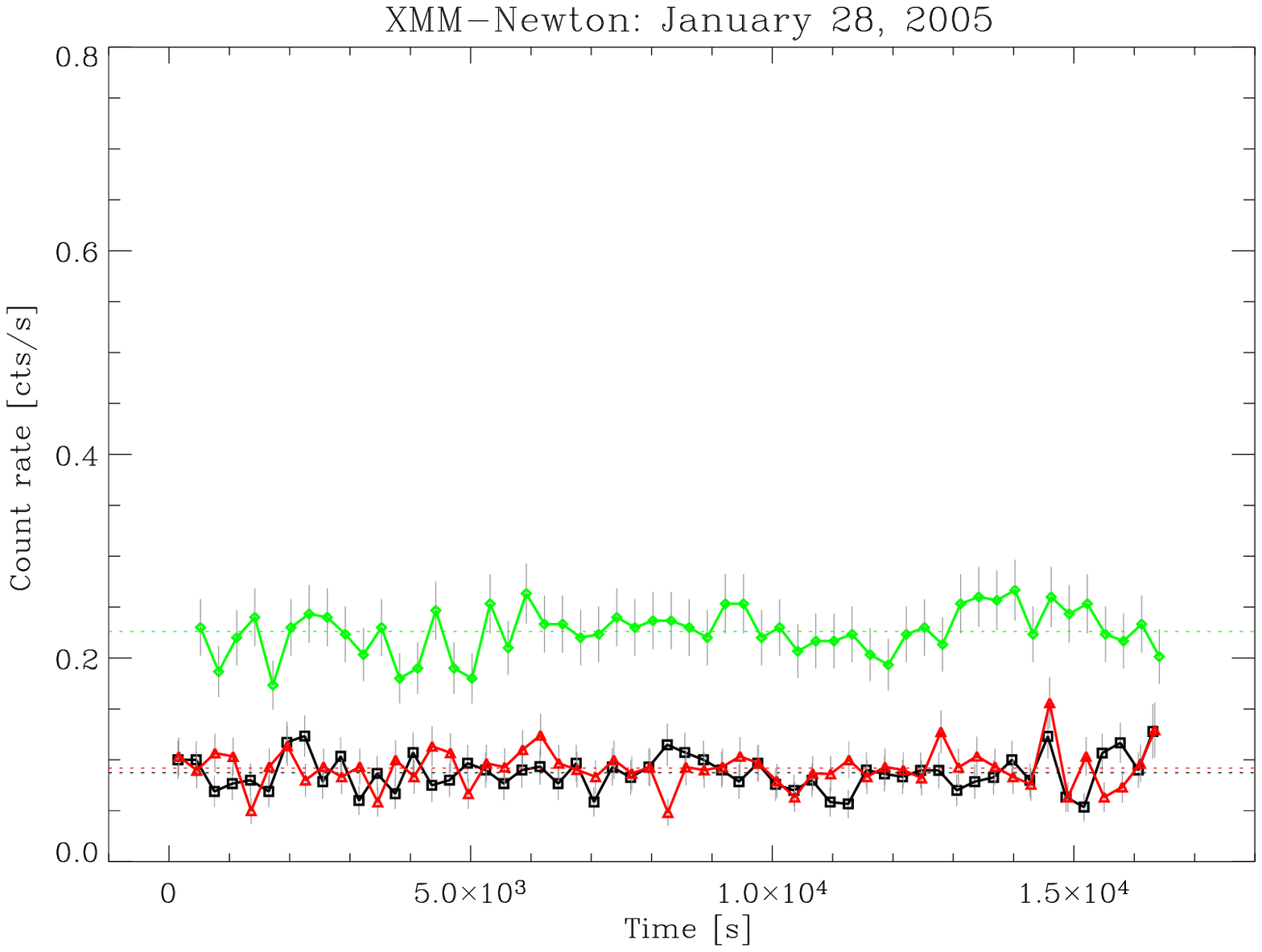}}
   \caption{X-ray light curves extracted from the XMM-Newton observation of January 28, 2005;
   data from pn (green diamonds), MOS1 (black squares), and MOS2 (red triangles) have been binned over 300 s;
   horizontal dotted lines indicate the average values of the datasets; vertical lines highlight the main peaks}
   \label{XMM3_lc}
   \end{figure}

Light curves of the source were extracted from the same regions where source spectra were taken,
and preliminarily binned in 1 s intervals. Only bins belonging to the good time intervals selected
during the filtering of high background periods were maintained and a second binning of 300 s
was performed in order to increase the signal to noise ratio.
These binned light curves are shown in Figs.\ \ref{Chandra_lc}--\ref{XMM3_lc}. For the XMM-Newton
observations black squares, red triangles,
and green diamonds refer to the MOS1, MOS2, and pn data, respectively. Dotted lines indicate the average values,
with the same colours used for the corresponding light curves.

In order to investigate the existence of a long-term trend and, at the same time, 
the presence of fast variations, we verify the goodness of a linear fit.
In Table \ref{lc} we report, for each observation and each instrument, the mean count rate $<C>$
of the binned light curve, its standard deviation $\sigma$, the parameters of 
a linear fit $y=a+b \, x$, the corresponding $\chi ^2 / \nu$ goodness-of-fit,
and the probability that the computed fit would have a value of $\chi ^2$ or greater.
If this probability is greater than 0.1, the fit is believable.
In the last column we also show the mean fractional variation, which is a
common parameter to characterize variability (Peterson \cite{pet01}):
$f_{\rm var}=\sqrt{\sigma^2-\delta^2}/<C>$, 
where $\delta^2$ is the mean square uncertainty of the count rates.

\begin{table*}
\centering
\caption{Statistics on the Chandra and XMM-Newton light curves binned every 300 s
and fitted with a straight line $y=a+b \, x$}
\begin{tabular}{l c c c c c c c}
\hline
Detector & $<C>$      & $\sigma$ &$a$&$b$ & $\chi ^2 / \nu$ & Prob.& $f_{\rm var}$\\
         &(cts/s)     &(cts/s)  &   &    &                 &      & \\
\hline
\hline
\multicolumn{8}{c}{Chandra: August 20--21, 2000}\\
\hline
ACIS-S     & 0.14 & 0.02 & 0.14 & $-5.36 \times 10^{-8}$ & 1.00 & 0.49 & 0.00\\
\hline
\hline
\multicolumn{8}{c}{XMM-Newton: February 10, 2002}\\
\hline
MOS1     & 1.08 & 0.15 & 1.20 & $-1.40 \times 10^{-5}$ &  4.26 & 0.00 & 0.13\\
MOS2     & 1.11 & 0.15 & 1.23 & $-1.40 \times 10^{-5}$ &  4.11 & 0.00 & 0.13\\
pn       & 3.36 & 0.38 & 3.62 & $-2.81 \times 10^{-5}$ & 10.52 & 0.00 & 0.11\\
\hline
\hline
\multicolumn{8}{c}{XMM-Newton: January 18--19, 2004}\\
\hline
MOS1     & 0.19 & 0.02 & 0.18 & $ 6.82 \times 10^{-7}$ &  0.82 & 0.89 & 0.00\\
MOS2     & 0.20 & 0.03 & 0.20 & $ 1.03 \times 10^{-8}$ &  1.01 & 0.46 & 0.00\\
pn       & 0.47 & 0.05 & 0.46 & $ 5.90 \times 10^{-7}$ &  1.15 & 0.16 & 0.03\\
\hline
\hline
\multicolumn{8}{c}{XMM-Newton: August 2, 2004}\\
\hline
MOS1     & 0.19 & 0.03 & 0.18 & $ 1.12 \times 10^{-6}$ &  1.01 & 0.46 & 0.04\\
MOS2     & 0.19 & 0.03 & 0.19 & $ 4.22 \times 10^{-7}$ &  0.98 & 0.51 & 0.00\\
pn       & 0.49 & 0.04 & 0.50 & $-1.31 \times 10^{-6}$ &  0.89 & 0.66 & 0.00\\
\hline
\hline
\multicolumn{8}{c}{XMM-Newton: January 28, 2005}\\
\hline
MOS1     & 0.09 & 0.02 & 0.08 & $-1.08 \times 10^{-7}$ &  1.05 & 0.38 & 0.04\\
MOS2     & 0.09 & 0.02 & 0.09 & $-4.88 \times 10^{-8}$ &  1.23 & 0.13 & 0.07\\
pn       & 0.22 & 0.02 & 0.21 & $ 1.43 \times 10^{-6}$ &  0.67 & 0.97 & 0.00\\
\hline
\label{lc}
\end{tabular}
\end{table*}

Figure \ref{XMM0_lc} shows the light curves derived from the XMM-Newton observation of
February 10, 2002.
As one can see, at least three episodes of remarkable variability can be recognized in the
light curves at the beginning (only MOS detectors), at about half exposure (especially in the pn detector,
at about $\sim 11000$ s), and at the end of the exposure. 
They imply variations of about 45\% in the count rates on time scales
$t=<C>/( | \Delta C/ \Delta t | ) \sim 40$ min.
The variability is confirmed by the high values of
$\chi ^2 / \nu$ and null probability reported in Table \ref{lc}. 
Notice also the decreasing trend highlighted by
the linear fit calculated on the pn data (solid green line in the figure). 
These results imply that there are both long-term and fast count rate changes.
The pn light curve of this XMM-Newton observation has been analyzed within several energy
bands between 0.3 keV and 12 keV by Kadler (\cite{kad05b}). No time lags between
the soft and hard energy bands could be detected down to a limit of $\sim$ 
200 s. However, a small but significant spectral steepening with
increasing flux is found from the extraction of separate spectra for the 
low and high states of the source emission.

On the contrary, no significant variability has been found when analysing the X-ray light
curves of the other pointings, which found the source in a much fainter state.
This is already visible in Figs.\ \ref{Chandra_lc} and \ref{XMM1_lc}--\ref{XMM3_lc}, 
since it is not possible to recognize variations which are detected by all instruments.
A confirmation comes from the statistical analysis whose results are shown in Table \ref{lc}:
the best-fit line is essentially a constant coincident
with the average value, indicating that there are no long-term trends.
Moreover, both the low $\chi ^2 / \nu$ values and the probability much greater than
0.1 indicate that this constant value is a good description of the data.
The picture is further confirmed by the mean fractional variation: only in the case
of the February 2002 observations  $f_{\rm var}$ is greater than 10\%.

\section{Conclusions}

We have analysed the data obtained during four XMM-Newton and one Chandra pointings of the blazar AO 0235+16
in the years 2000--2005. 
This analysis has carefully taken into account the
contribution to both emission and absorption from the complex environment at redshift $z=0.524$.

We have fitted the X-ray spectra with different models (single power law, logarithmic parabola, broken power law,
curved model, double power law) and have found that those implying curvature seem to give a better
description of the source spectra.
The double power law implies pronounced spectral curvature at soft X-rays.
Such curvature is not implausible, and favoured in this paper, with regard to the expected
overlapping of the synchrotron (or thermal) and inverse-Compton branches of the SED in the energy range
discussed here (0.3--10 keV).

We have speculated about the possible existence of a Fe K$\alpha$ emission feature at $\sim 2.8$ keV
in both the Chandra and August 2004 XMM-Newton spectra.
If this tentative detection is real, it may suggest strong gravitational redshift effects
from the inner accretion disc.

The optical--UV data from the Optical Monitor onboard XMM-Newton allow one to reconstruct the optical to X-ray
SED with simultaneous data.
The optical--UV brightness level is higher when the X-ray one is higher.
The SED reveals a UV--soft-X-ray bump, which was previously noticed by Junkkarinen et al.\ (\cite{jun04})
and Raiteri et al.\ (\cite{rai05}).
If this bump is intrinsic to the source, one possible mechanism
producing this extra-component may be synchrotron emission from an inner region of the jet
with respect to the region where the lower-energy synchrotron component comes from.
However, thermal emission from an accretion disc,
similar to the ``big blue bump" observed in many AGNs, cannot be ruled out.
Investigating whether this extra-component
is a common feature of the blazar class and
discovering the mechanism which is responsible for it are two intriguing issues which
certainly deserve further multiwavelength observing efforts.

We have detected remarkable short-term variability in the X-ray light curves derived from the XMM-Newton observation
of February 10, 2002, when the source was in a bright state. 
We noticed three episodes where the count rate changed by about 45\% on a time scale of about 40 min.
On the contrary, no significant variability was evident
in the X-ray light curves corresponding to the other observing epochs analysed in this paper.

\begin{acknowledgements}
C. M. R. is grateful to L.\ Foschini and A.\ Capetti for useful discussions and to B.\ Balmaverde for a help with CIAO.
M. K. was supported for this research by the
International Max Planck Research School for Radio and Infrared Astronomy, 
funded by the Max-Planck-Gesellschaft.
This work was partly supported by the European Community's Human Potential Programme
under contract HPRN-CT-2002-00321 (ENIGMA)
and by the Italian MIUR under grant Cofin 2003/2003027534\_002.
\end{acknowledgements}

\end{document}